\documentclass[10pt]{amsart}
\usepackage{amsmath,amssymb,amscd}
\usepackage[dvips]{epsfig}

\newtheorem{theorem}{Theorem}
\newtheorem{lemma}[theorem]{Lemma}
\newtheorem{corollary}[theorem]{Corollary}
\newtheorem{proposition}[theorem]{Proposition}
\newtheorem{definition}{Definition}

\newtheorem{conjecture}[theorem]{Conjecture}

\allowdisplaybreaks \numberwithin{equation}{section}

\newcommand{\PP}{{\mathbb P}}

\def\Jac{\mathop{\rm Jac}\nolimits}

\def\dim{\mathop{\rm dim}\nolimits}

\begin{document}

\title[On the tetrahedrally symmetric monopole
]{On the tetrahedrally symmetric monopole}
\author{H.W. Braden}
\address{School of Mathematics, Edinburgh University, Edinburgh.}
\email{hwb@ed.ac.uk}
\author{V.Z. Enolski}
\address{Institute of Magnetism, National Academy of Sciences of
Ukraine.} \email{vze@ma.hw.ac.uk}

\begin{abstract}
We study $SU(2)$ BPS monopoles with spectral curves of the form
$\eta^3+\chi(\zeta^6+b \zeta^3-1)=0$. Previous work has has
established a countable family of solutions to Hitchin's
constraint that $L^2$ was trivial on such a curve. Here we
establish that the only curves of this family that yield BPS
monopoles correspond to tetrahedrally symmetric monopoles. We
introduce several new techniques making use of a factorization
theorem of Fay and Accola for theta functions, together with
properties of the Humbert variety. The geometry leads us to a
formulation purely in terms of elliptic functions. A more general
conjecture than needed for the stated result is given.
\end{abstract}

\maketitle

\tableofcontents

\section{Introduction}

For many years there has been considerable interest in magnetic
monopoles, the topological soliton solutions of Yang-Mills-Higgs
gauge theories in three space dimensions with particle-like
properties. In particular BPS monopoles have been the focus of
much research (see \cite{ms04} for a recent review). These
monopoles satisfy a rather ubiquitous first order Bogomolny
equation
$$B_i=\frac{1}{2}\sum_{j,k=1}\sp3\epsilon_{ijk}F\sp{jk}=D_i\Phi$$
(together with certain boundary conditions, the remnant of a limit
in which the Higgs potential is removed). Here $F_{ij}$ is the
field strength associated to a gauge field $A$, and $\Phi$ is the
Higgs field.  The Bogomolny equation may be viewed as a
dimensional reduction of the four dimensional self-dual equations
upon setting all functions independent of $x_4$ and identifying
$\Phi=A_4$; they are also encountered in supersymmetric theories
when requiring certain field configurations to preserve some
fraction of supersymmetry. The study of BPS monopoles is
intimately connected with integrable systems. Nahm gave a
transform of the ADHM instanton construction to produce BPS
monopoles \cite{nahm82} and the resulting Nahm's equations have
Lax form with corresponding spectral curve $\hat{\mathcal{C}}$.
Hitchin gave a twistorial description of this curve
\cite{hitchin82}: just as Ward's twistor transform relates
instanton solutions in $\mathbb{R}\sp4$ to certain holomorphic
vector bundles over the twistor space $\mathbb{CP}\sp3$, Hitchin
showed that the dimensional reduction leading to BPS monopoles
could be made at the twistor level as well and also obtained the
same curve lying in mini-twistor space, $\hat{\mathcal{C}}\subset$
T$\mathbb{P}\sp1$. Subject to certain nonsingularity conditions on
the curve $\hat{\mathcal{C}}$ Hitchin was able to prove all
monopoles could be obtained by this approach \cite{hitchin83}.
Bringing methods from integrable systems to bear upon the
construction of solutions to Nahm's equations for the gauge group
$SU(2)$ Ercolani and Sinha \cite{es89} later showed how one could
solve (a gauge transform of) the Nahm equations in terms of a
Baker-Akhiezer function for the curve $\hat{\mathcal{C}}$.

Despite the many striking results now obtained, disappointingly
few explicit solutions are known. This paper is part of a longer
reappraisal of this work, seeking to understand where the
difficulties in implementation lie and developing new techniques
to address them. Throughout we shall focus on $SU(2)$ BPS
monopoles and this paper will deal with curves $\hat{\mathcal{C}}$
of the form
\begin{equation} \eta^3+\chi(\zeta^6+b
\zeta^3-1)=0\label{bren03}
\end{equation}
where $\chi$, $b$ are certain real parameters. Such a curve is of
genus $4$ and could describe a charge three monopole. Two types of
problem arise (that will be made more precise below). The first is
that the curve (\ref{bren03}) is subject to a set of constraints
whereby  the periods of a meromorphic differential on the curve
are specified. This type of constraint arises in many other
settings as well, for example when specifying the filling
fractions of a curve in the AdS/CFT correspondence. Such
constraints are transcendental in nature and the number of cases
where they have be solved explicitly is rather limited. This is
certainly an area that needs to be studied more. For the curve
(\ref{bren03}) a countable number of solutions to this constraint
have been obtained. The second type of problem arises in
implementing a constraint that may be expressed as the vanishing
of a real one parameter family of cohomologies of certain line
bundles, $H^0(\hat{\mathcal{C}},L^{\lambda}(n-2))=0$ for
$\lambda\in(0,2)$. Viewing the line bundles as points on the
Jacobian this is equivalent to a real line segment not
intersecting the theta divisor $\Theta$ of the curve. Indeed there
are sections for $\lambda=0,2$ and the flow is periodic (mod $2$)
in $\lambda$ and so we are interested in the number of times a
real line intersects $\Theta$. While techniques exist that count
the number of intersections of a complex line with the theta
divisor we are unaware of anything comparable in the real setting.
We establish here that of the countable number of solutions to the
second constraint that
\begin{theorem}\label{mainthm}
The only curves of the family (\ref{bren03}) that yield BPS
monopoles correspond to tetrahedrally symmetric monopoles. These
have $b=\pm5\sqrt{2}$, $\chi^{\frac{1}{3}} =-\frac16\,
    \frac{\Gamma(\frac16)\Gamma(\frac13)}{2\sp\frac16\,
    \pi\sp{\frac12}}$.
\end{theorem}

An outline of the paper is as follows. In section 2 we will recall
the constraints on the curve $\hat{\mathcal{C}}$  that are
equivalent for a monopole to exist. We shall then describe the
curve (\ref{bren03}) in more detail and make concrete these
constraints for this curve. This will entail a description of the
homology and period matrix for the curve. At this stage we will
have reduced the problem to properties of the theta function for
the genus $4$ curve $\hat{\mathcal{C}}$. (Our theta function
conventions are given in Appendix A.) Now the curve (\ref{bren03})
(and indeed the more general curve
$\eta^3+\alpha\eta\zeta^2+\chi(\zeta^6+b \zeta^3-1)=0$ that will
be explored elsewhere) is invariant under the cyclic group
$\texttt{C}_3$ and we have a covering
$\pi:\hat{\mathcal{C}}\rightarrow\mathcal{C}$ of a genus $2$ curve
$\mathcal{C}$. In section 3 we will describe a theorem of Fay and
Accola that allows us to express the genus 4 theta functions in
terms of genus 2 theta functions so reducing the problem to one of
genus 2 theta functions. Then in section 4 we will use the
reduction theory of Humbert to further reduce the problem to that
of elliptic functions. At this stage we have reduced the initial
problem of the existence of a BPS monopole to a question about the
number of zeros of an elliptic function on an interval. Section 5
describes this in more detail. Although we have a stronger
conjecture than we can prove, we are able to establish the theorem
given above. We end with some final observations in section 6.

\section{The Spectral Curve and its Constraints}

\subsection{Hitchin Data}
If $\zeta$ is the inhomogeneous coordinate on the Riemann sphere,
and $(\zeta,\eta)$ are the standard local coordinates on
$T\PP\sp1$ (defined by
$(\zeta,\eta)\rightarrow\eta\frac{d}{d\zeta}$), the spectral curve
of a charge $n$ monopole $\hat{\mathcal{C}}\subset$
T$\mathbb{P}\sp1$ may be expressed in the form
\begin{equation}
P(\eta,\zeta)=\eta^n+\eta^{n-1} a_1(\zeta)+\ldots+\eta^r
a_{n-r}(\zeta)+ \ldots+\eta\,
a_{n-1}(\zeta)+a_n(\zeta)=0,\label{spectcurve}
\end{equation}
where $a_r(\zeta)$  (for $1\leq r\leq n$) is a polynomial in
$\zeta$ of maximum degree $2r$. We may view $\hat{\mathcal{C}}$ as
an $n$-fold branched cover of $\mathbb{P}\sp{1}$ and (by a
rotation if necessary) we may assume $n$ distinct preimages
$\{\infty_k\}_{k=1}\sp{n}$ of the point $\zeta=\infty$. The form
of the curve means that $\eta/\zeta\sim \rho_k \zeta$ at
$\infty_k$. For a generic $n$-monopole the spectral curve is
irreducible and has genus $g_{\hat{\mathcal{C}}}=(n-1)^2$. We will
denote by
$\{\hat{\mathfrak{a}}_k,\hat{\mathfrak{b}}_k\}_{k=1}\sp{g_{\hat{\mathcal{C}}}}$
a canonical homology basis of $\hat{\mathcal{C}}$.

The Hitchin data constrains the curve $\hat{\mathcal{C}}$
explicitly in terms of the polynomial $P(\eta,\zeta)$ and
implicitly in terms of the behaviour of various line bundles on
$\hat{\mathcal{C}}$. If the homogeneous coordinates of $ \PP\sp1$
are $[\zeta_0,\zeta_1]$ we consider the standard covering of this
by the open sets $U_0=\{[\zeta_0,\zeta_1]\,|\,\zeta_0\ne0\}$ and
$U_1=\{[\zeta_0,\zeta_1]\,|\,\zeta_1\ne0\}$, with
$\zeta=\zeta_1/\zeta_0$ the usual coordinate on $U_0$. We will
denote by $\hat U_{0,1}$ the pre-images of these sets under the
projection map $\pi:T\PP\sp1\rightarrow\PP\sp1$. Let
 $L^{\lambda}$ denote the holomorphic line bundle on
$T\PP\sp1$ defined by the transition function
$g_{01}=\rm{exp}(-\lambda\eta/\zeta)$ on $\hat U_{0}\cap \hat
U_{1}$, and let $L^{\lambda}(m)\equiv
L^{\lambda}\otimes\pi\sp*\mathcal{O}(m)$ be similarly defined in
terms of the transition function
$g_{01}=\zeta^m\exp{(-\lambda\eta/\zeta)}$. A holomorphic section
of such line bundles is given in terms of holomorphic functions
$f_\alpha$ on $\hat U_\alpha$ satisfying
$f_\alpha=g_{\alpha\beta}f_\beta$. We denote line bundles on
$\mathcal{C}$ in the same way, where now we have holomorphic
functions $f_\alpha$ defined on $\mathcal{C}\cap\hat U_\alpha$.

The Hitchin data constrains the curve to satisfy:\\
\begin{description}
\item[H1] The curve $\hat{\mathcal{C}}$ is real with respect
    to the standard real structure on $T\PP\sp1$ (the
    anti-holomorphic involution defined by reversing the
orientation of the lines in ${\mathbb R}\sp3$),
\begin{equation}
\tau:(\zeta,\eta)\mapsto(-\frac{1}{\bar{\zeta}},
-\frac{\bar{\eta}}{\bar{\zeta}^2}).
\end{equation}

\item[H2] $L^2$ is trivial on $\hat{\mathcal{C}}$ and $L(n-1)$
    is real.

\item[H3] $H^0(\hat{\mathcal{C}},L^{\lambda}(n-2))=0$ for $\lambda\in(0,2)$.\\
\end{description}

Only the first of these constraints is immediate to implement. The
reality of the curve means the coefficients of (\ref{spectcurve})
satisfy
\begin{equation}\label{spectcurvereal}
a_r(\zeta)=(-1)^r\zeta^{2r}\overline{a_r(-\frac{1}{\overline{\zeta}})}
.\end{equation} For the curve (\ref{bren03}) this is why $\chi$
and $b$ are real. Ercolani and Sinha show the reality of $L(n-1)$
within the Baker-Akhiezer function setting and \cite{bren06}
implements this in terms of theta functions on the curve. The
triviality of $L^2$ means that there exists a nowhere-vanishing
holomorphic section. In terms of the open sets $\hat U_{0,1}$ we
will have two, nowhere-vanishing holomorphic functions, $f_0$ on
$\hat U_0\cap\hat{\mathcal{C}}$ and $f_1$ on $\hat
U_1\cap\hat{\mathcal{C}}$, such that on the intersection
\begin{equation}
 f_{0}(\eta,\zeta)=\mathrm{exp} \left\{
-2\frac{\eta}{\zeta} \right\} f_1(\eta,\zeta). \label{triv3}
\end{equation}
Consideration of the logarithmic derivative of (\ref{triv3}) shows
that, in order to avoid essential singularities, we must have
\begin{align}
\mathrm{d}\mathrm{log}\,f_{0}(P) &=\left(\frac{2\rho_k}{t^2}
+O(1)\right){d}t,\quad \text{at}\quad P\rightarrow
\infty_k,\label{foex}
\end{align}
where $t=1/\zeta$ is a local parameter. Ercolani and Sinha
introduced the normalized meromorphic differential $\gamma_\infty$
whose pole behaviour is that of
$\frac12\mathrm{d}\mathrm{log}\,f_{0}(P)$ and whose
$\mathfrak{a}$-periods vanish. In terms of the vector of
$\mathfrak{b}$-periods we have
\begin{lemma}[Ercolani-Sinha Constraints]\label{EScond} The following are equivalent:
\begin{enumerate} \item $L\sp2$ is trivial on $\hat{\mathcal{C}}$.

\item $2\widehat{\boldsymbol{U}}\in \Lambda\Longleftrightarrow
    \widehat{\boldsymbol{U}}=\frac{1}{2\pi\imath}\left(\oint_{\hat{\mathfrak{b}}_1}\gamma_{\infty},
    \ldots,\oint_{\hat{\mathfrak{b}}_{g_{\hat{\mathcal{C}}}}}\gamma_{\infty}\right)=
    \frac12 \boldsymbol{n}+\frac12\hat\tau\boldsymbol{m}  $
    where $\boldsymbol{n}$,
    $\boldsymbol{m}\in\mathbb{Z}\sp{\hat g}$.

\item There exists a 1-cycle
    $\mathfrak{c}=\boldsymbol{n}\cdot{\hat{\mathfrak{a}}}+
    \boldsymbol{m}\cdot{\hat{\mathfrak{b}}}$ such that
    $\oint\limits_{\mathfrak{c}}\Omega=-2\beta_0$ for every
    holomorphic differential
    $\Omega=\dfrac{\beta_0\eta^{n-2}+\beta_1(\zeta)\eta^{n
-3}+\ldots+\beta_{n-2}(\zeta)}{\frac{\partial{P}}{\partial
\eta}}\,d\zeta$, where $\beta_j(\zeta)$ is a polynomial of
degree at most $2j$ in $\zeta$.

\end{enumerate}
\end{lemma}
Here $\Lambda$ is the period lattice of $\hat{\mathcal{C}}$ and
$\hat\tau$ the $\mathfrak{a}$-normalized period matrix. Ercolani
and Sinha established the equivalence of (1) and (2) while  the
dual form of the Ercolani-Sinha constraints (3) was given by
Houghton, Manton and Ram\~ao \cite{hmr99}. If the anti-holomorphic
involution $\tau$ induces an action $M_\tau$ on the homology,
$\begin{pmatrix}\tau_\ast \hat{\mathfrak{a}}\\ \tau_\ast
\hat{\mathfrak{b}}\end{pmatrix}=M_\tau \begin{pmatrix} \hat{\mathfrak{a}}\\
\hat{\mathfrak{b}}\end{pmatrix}$, then we have $M_\tau
JM_\tau=-J$, where $J$ is the standard symplectic structure, and
\begin{corollary}\label{HMRinvc}
$\tau_*\mathfrak{c}=-\mathfrak{c}$ or
$2\widehat{\mathbf{U}}M_\tau=
\begin{pmatrix}
  \mathbf{n} & \mathbf{m}
\end{pmatrix}
M_\tau=-
\begin{pmatrix}
  \mathbf{n} & \mathbf{m} \end{pmatrix}
$.
\end{corollary}
The Ercolani-Sinha constraints place $g_{\hat {\mathcal{C}}}$
transcendental constraints on the spectral curve $
\hat{\mathcal{C}}$ and, as noted in the introduction,  solving
these is a major difficulty in implementing this theory.

We have yet to discuss \textbf{H3}, the implementation of which is
the second type of problem mentioned in the introduction. Here
$L^{\lambda}(n-2)$ is a degree $g_{\hat {\mathcal{C}}} -1$ line
bundle so using Riemann's vanishing theorem for line bundles
$\mathcal{L}$ of this degree, ${\rm multiplicity}_\mathcal{L}\,
\theta =\dim H\sp0(\mathcal{C},\mathcal{O}(\mathcal{L}))$, we see
that $L^{\lambda}(n-2)$ does not lie in the theta divisor for
$\lambda\in(0,2)$. In \cite{bren06} we establish that
\begin{lemma} Let $\widetilde{\boldsymbol{K}}=
\boldsymbol{K}+\boldsymbol{\phi}\left((n-2)
\sum_{k=1}\sp{n}\infty_k\right)$ where $\boldsymbol{K}$ is the
vector of Riemann constants and $\boldsymbol{\phi}$ the
Abel-Jacobi map. Then
\begin{equation}\label{beh3}
H^0(\hat{\mathcal{C}},L^{\lambda}(n-2))\ne0\Longleftrightarrow
\theta(\lambda\widehat{\boldsymbol{U}}-
\widetilde{\boldsymbol{K}}\,|\,\hat\tau)=0
\end{equation} for $\lambda\in(0,2)$,
where $\theta$ is Riemann's theta function for the curve
$\hat{\mathcal{C}}$.
\end{lemma}
Thus the second problem is to determine when the (real) line
$\lambda\widehat{\boldsymbol{U}}- \widetilde{\boldsymbol{K}}$
intersects the theta divisor $\Theta$. We note the following
properties of the vector $\widetilde{\boldsymbol{K}}$,
\begin{lemma}
\hfill
\begin{itemize}
\item $\widetilde{\boldsymbol{K}}$ is independent of the
    choice of base point of the Abel map.
\item $2\widetilde{\boldsymbol{K}}\in\Lambda$.
\item For $n\ge3$ then $\widetilde{\boldsymbol{K}}\in
    \Theta_{\rm singular}$, the singular locus of the theta
    divisor.
\item $\widehat{\boldsymbol{U}}\pm \widetilde{\boldsymbol{K}}$
    is a non-singular even theta characteristic.
\end{itemize}
\end{lemma}
It is straightforward to see that $2\widehat{\boldsymbol{U}}\ne0$
and is a primitive vector in the period lattice. We also remark
that because $\widetilde{\boldsymbol{K}}$ is a half-period then
\begin{equation}\label{beh3b}
\theta(\lambda\widehat{\boldsymbol{U}}-
\widetilde{\boldsymbol{K}}\,|\,\hat\tau)=0
\Longleftrightarrow
\theta(\lambda\widehat{\boldsymbol{U}}+
\widetilde{\boldsymbol{K}}\,|\,\hat\tau)=0.
\end{equation}

\subsection{The curve and its properties}
We will write the curve (\ref{bren03}) in the form
\begin{equation}
w^3=z^6+bz^3-1=(z^3-\alpha\sp3)(z^3+\frac{1}{\alpha\sp3})\label{cyccurve}
\end{equation}
to avoid various factors, where $(z,w)=(\zeta,-\eta/\chi^{1/3})$
and $1/\alpha\sp3=(b+\sqrt{b^2+4})/{2}$. With
$\rho=e\sp{2\imath\pi /3}$ this curve has symmetries:
$$\mathrm{R}:\ (z,w)\rightarrow  (z,\rho{ w}),\qquad \sigma:\
(z,w)\rightarrow (\rho z, {\rho  w}),\qquad \mathrm{c}:\
(z,w)\rightarrow  (-1/z,-w/z^2).
$$
These yield the group $C_3\times S_3$, with
$C_3=<\mathrm{R}|\,\mathrm{R}^3=1>$ and
$S_3=<\sigma,\mathrm{c}|\,\sigma^3=1,\mathrm{c}^2=1,\mathrm{c}\sigma
\mathrm{c}=\sigma^2>$. When $b=\pm5\sqrt{2}$, the dihedral
symmetry $S_3$ is enlarged to tetrahedral symmetry by
$$ \mathrm{t}:\ (z,w)\rightarrow \left(\pm\,\frac{\sqrt{2}\mp z}{1\pm\sqrt{2}z},
-\frac{3  w}{(1\pm\sqrt{2}z)^2}\right),\qquad \mathrm{t}^2=1,$$ with $A_4$
being generated by $\sigma$ and $\mathrm{t}$.

\begin{figure}[scale=1.25]
\setlength{\unitlength}{\textwidth}
\begin{center}
\begin{picture}(1,1)
\put(0.5,0.6){\includegraphics[width=0.5\unitlength]{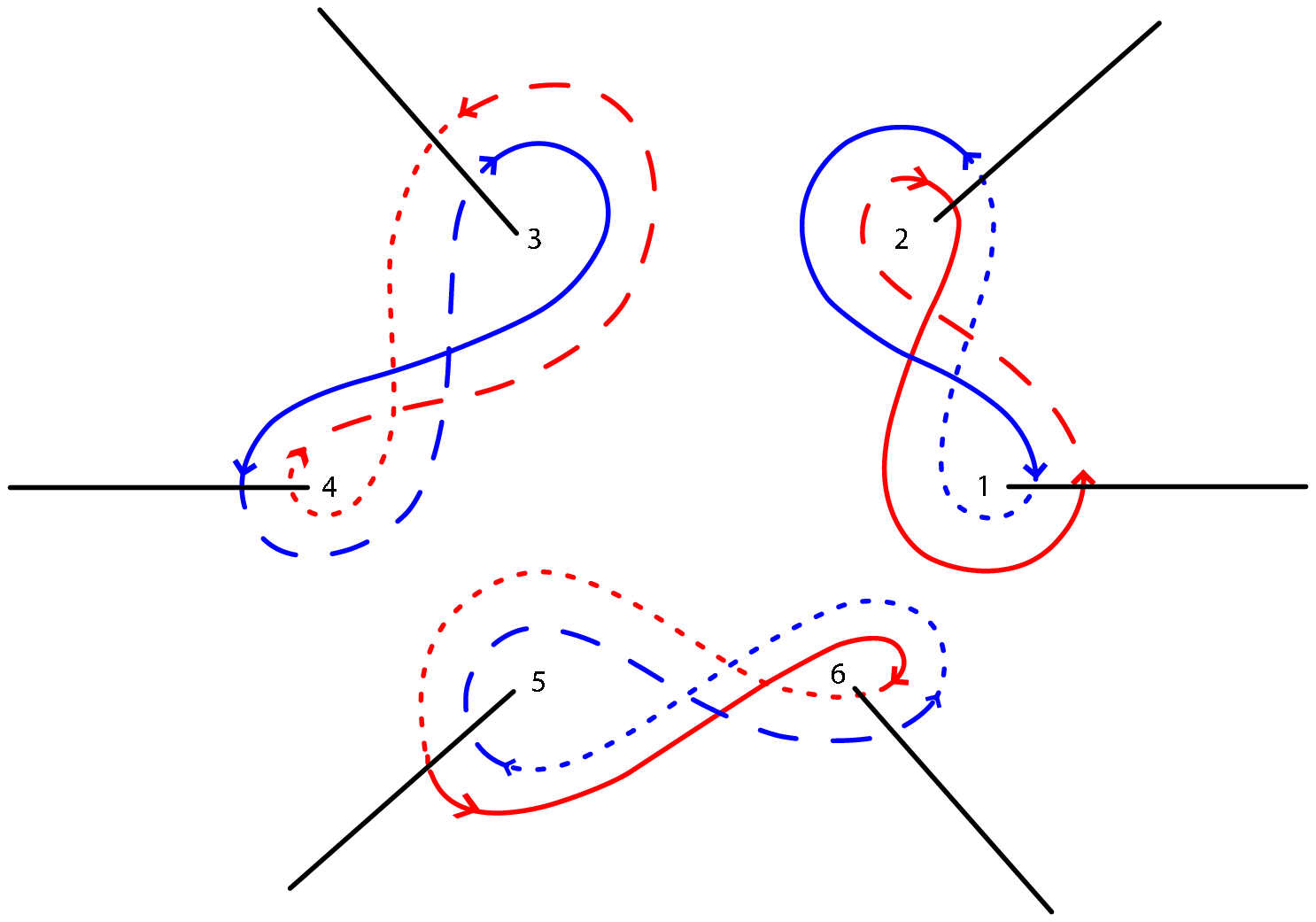}}
\put(0.5,0.14){\includegraphics[width=0.5\unitlength]{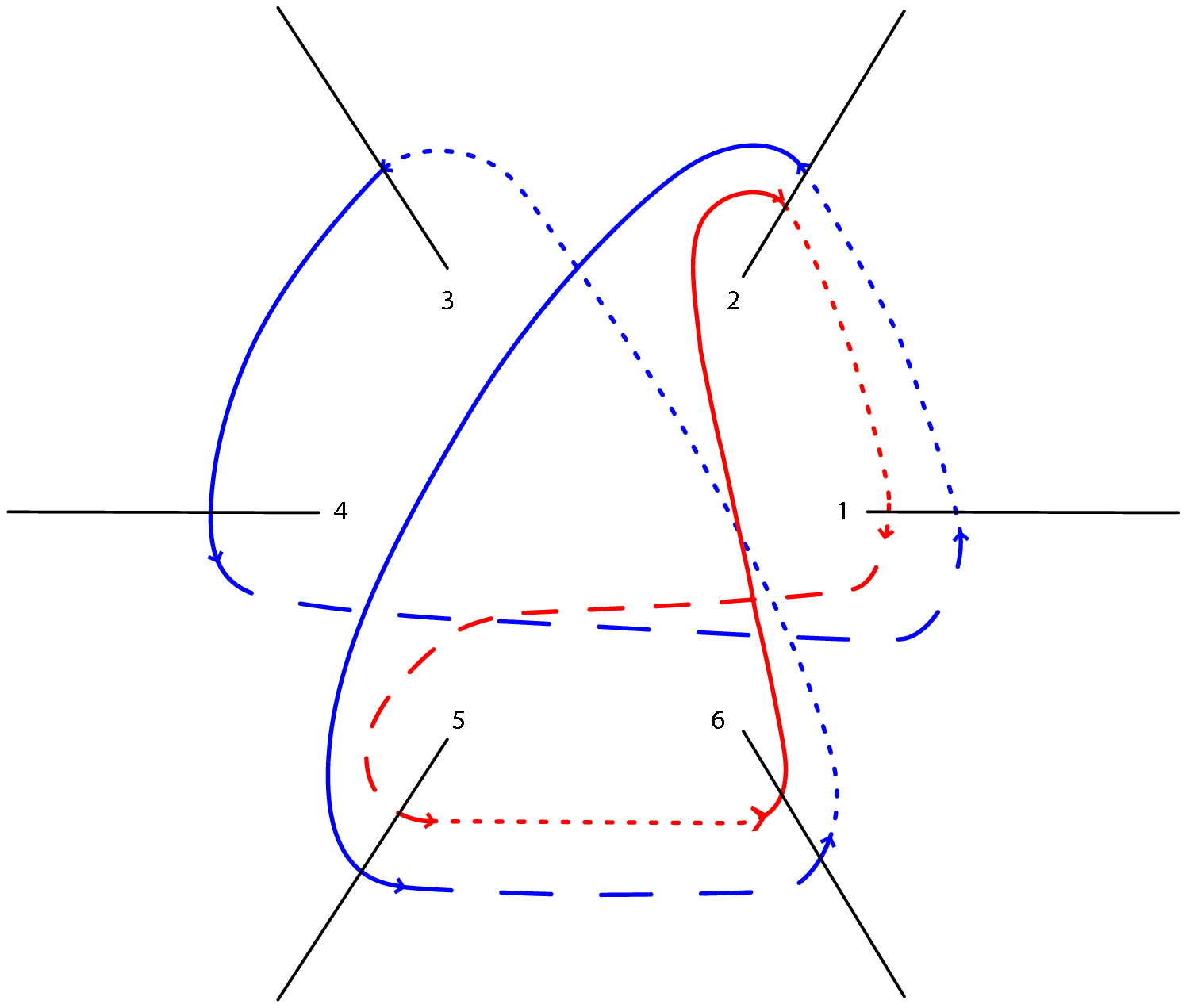}}
\put(0,0.6){\includegraphics[width=0.5\unitlength]{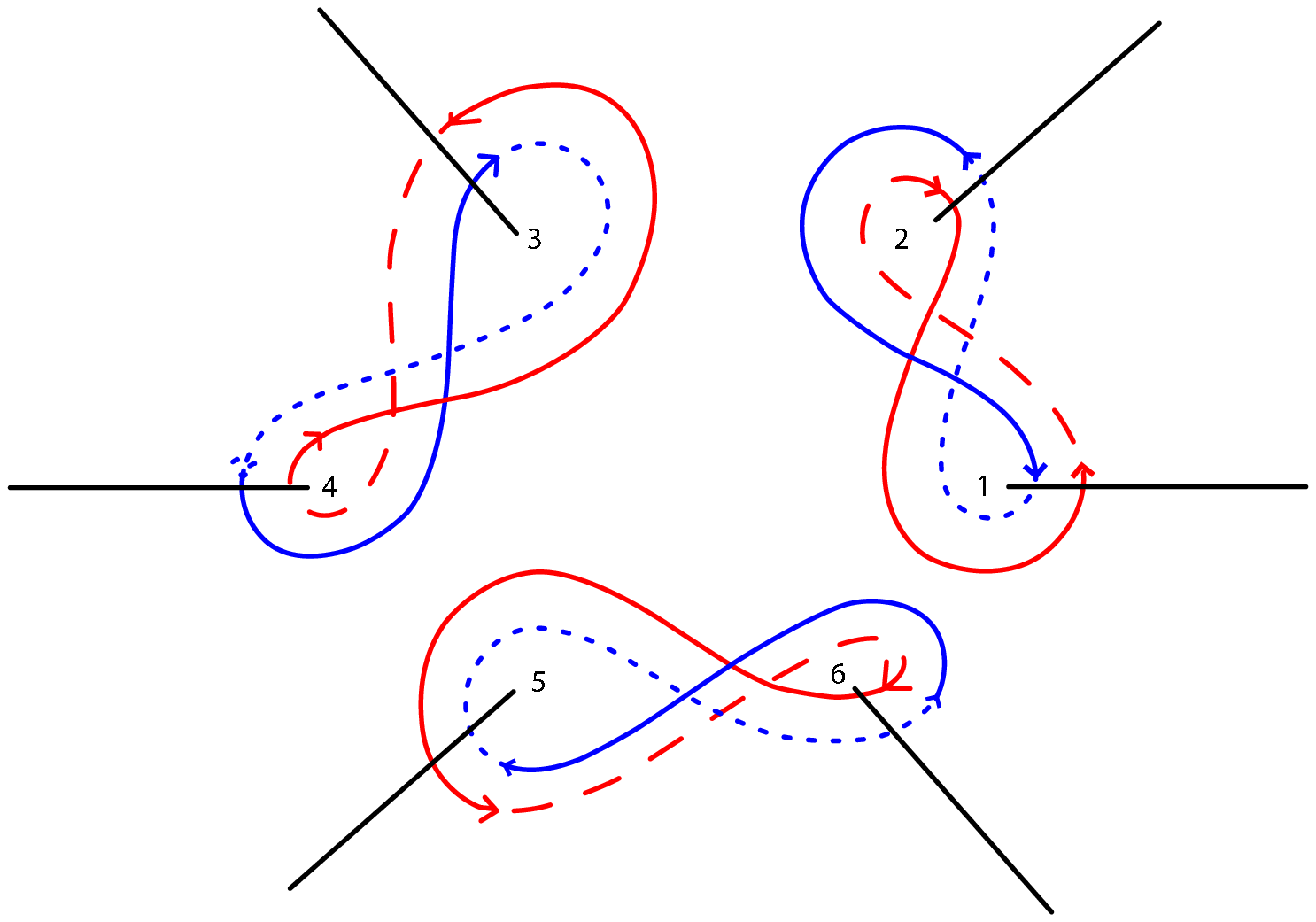}}
\put(0,0.14){\includegraphics[width=0.5\unitlength]{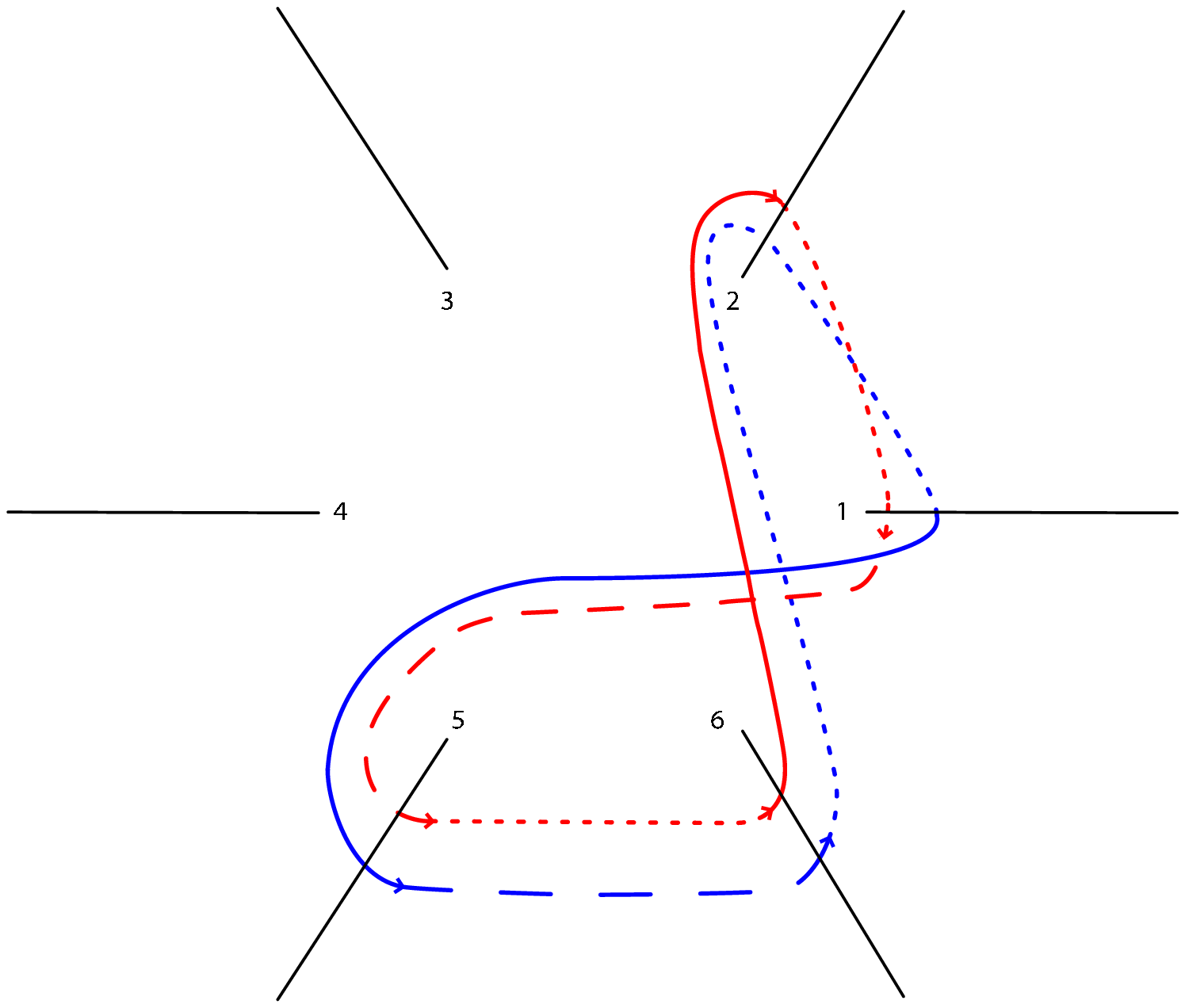}}
\put(0.1,0.1){(a) The symmetric basis $\hat{\mathfrak{a}}\sp{s}_\ast, \hat{\mathfrak{b}}\sp{s}_\ast$}
\put(0.6,0.1){(b) The cyclic basis $\hat{\mathfrak{a}}\sp{c}_\ast, \hat{\mathfrak{b}}\sp{c}_\ast$}
\put(0.37,0.71){$\hat{\mathfrak{a}}_1\sp{s}$}
\put(0.3,0.81){$\hat{\mathfrak{b}}_1\sp{s}$}
\put(0.2,0.79){$\hat{\mathfrak{a}}_2\sp{s}$}
\put(0.1,0.71){$\hat{\mathfrak{b}}_2\sp{s}$}
\put(0.17,0.61){$\hat{\mathfrak{a}}_3\sp{s}$}
\put(0.36,0.67){$\hat{\mathfrak{b}}_3\sp{s}$}
\put(0.87,0.71){$\hat{\mathfrak{a}}_1\sp{c}$}
\put(0.8,0.81){$\hat{\mathfrak{b}}_1\sp{c}$}
\put(0.7,0.79){$\hat{\mathfrak{a}}_2\sp{c}$}
\put(0.6,0.71){$\hat{\mathfrak{b}}_2\sp{c}$}
\put(0.67,0.61){$\hat{\mathfrak{a}}_3\sp{c}$}
\put(0.86,0.67){$\hat{\mathfrak{b}}_3\sp{c}$}
\put(0.235,0.225){$\hat{\mathfrak{a}}_4\sp{s}$}
\put(0.11,0.225){$\hat{\mathfrak{b}}_4\sp{s}$}
\put(0.735,0.225){$\hat{\mathfrak{a}}_0\sp{c}$}
\put(0.61,0.225){$\hat{\mathfrak{b}}_0\sp{c}$}
\end{picture}
\end{center}
\caption{The homology bases. Sheet 1 is denoted by a solid line; sheet 2 by a dashed line;
and sheet 3 by a dotted line.}
\label{fighom}
\end{figure}

The paper \cite{bren06} chose an homology basis\footnote{The
conventions of the paper \cite{bren06} were such that the
$\mathfrak{b}$-normalized period matrix was positive definite. We
must change the relative orientation of the $\mathfrak{a}$-cycles
and $\mathfrak{b}$-cycles to obtain the positive definite
$\mathfrak{a}$-normalized period matrix used in this paper.} $\{
\hat{\mathfrak{a}}_i\sp{s},\hat{\mathfrak{b}}_i\sp{s}\}$
reflecting the symmetry $\mathrm{R}$:
$\mathrm{R}(\hat{\mathfrak{b}}_i\sp{s})=-\hat{\mathfrak{a}}_i\sp{s}$
($i=1,2,3$) and
$\mathrm{R}(\hat{\mathfrak{b}}_4\sp{s})=\hat{\mathfrak{a}}_4\sp{s}$.
If we order the branch points $\{\lambda_1,\ldots,\lambda_6\}$ in
terms of increasing argument and denote by $\gamma_k(z_i,z_j)$ the
oriented path going  from $P_i=(z_i,w_i)$ to $P_j=(z_j,w_j)$ in
the  $k$-th sheet we may express these cycles as
\begin{align}\begin{split}
\hat{\mathfrak{a}}\sp{s}_1&=\gamma_1(\lambda_2,\lambda_1)+\gamma_2(\lambda_1,\lambda_2),
\qquad
\hat{\mathfrak{b}}\sp{s}_1=\gamma_1(\lambda_2,\lambda_1)+\gamma_3(\lambda_1,\lambda_2),\\
\hat{\mathfrak{a}}\sp{s}_2&=\gamma_1(\lambda_4,\lambda_3)+\gamma_2(\lambda_3,\lambda_4)
,\qquad
\hat{\mathfrak{b}}\sp{s}_2=\gamma_1(\lambda_4,\lambda_3)+\gamma_3(\lambda_3,\lambda_4),\\
\hat{\mathfrak{a}}\sp{s}_3&=\gamma_1(\lambda_6,\lambda_5)
+\gamma_2(\lambda_5,\lambda_6),\qquad
\hat{\mathfrak{b}}\sp{s}_3=\gamma_1(\lambda_6,\lambda_5)
+\gamma_3(\lambda_5,\lambda_6),\\
\hat{\mathfrak{a}}\sp{s}_4&=\gamma_3(\lambda_2,\lambda_1)
+\gamma_2(\lambda_1,\lambda_5)+\gamma_3(\lambda_5,\lambda_6)
+\gamma_1(\lambda_6,\lambda_2),
\\
\hat{\mathfrak{b}}\sp{s}_4&
=\gamma_2(\lambda_2,\lambda_1)+\gamma_3(\lambda_6,\lambda_2)
+\gamma_2(\lambda_5,\lambda_6)+\gamma_1(\lambda_1,\lambda_5).
\end{split}\label{homology_s}
\end{align}
This is shown in Figure 1(a). In the next section we shall choose
a homology basis reflecting the symmetry $\sigma$ in order to use
the results of Fay and Accola. Let us fix the following
lexicographical ordering of independent canonical holomorphic
differentials of $\hat{\mathcal{C}}$,
\begin{equation} {d}u_1= \frac{{d} z}{w},\quad
                 {d}u_2= \frac{{d} z}{w^2},\quad
                  {d}u_3= \frac{z{d} z}{w^2},\quad
                  {d}u_4= \frac{z^2{d} z}{w^2}.
\label{diffbasis}
 \end{equation}
 Then the symmetry $\mathrm{R}$ together with the Riemann
 bilinear relations shows that the period matrix for $\hat{\mathcal{C}}$
 may be expressed in terms of just the four periods
 $$\boldsymbol{x}=(x_1,x_2,x_3,x_4)^T
=\left(
\oint_{\mathfrak{a}_1}{d}u_1,\ldots,\oint_{\mathfrak{a}_4}{d}u_1
\right)^T.$$ Following Wellstein \cite{wel99} and Matsumoto \cite{matsu00} we find
\begin{proposition}
\label{matsumoto1} Let $\hat{\mathcal{C}}$ be the triple covering
of $\mathbb{P}^1$ with six distinct point $\lambda_1,
\ldots,\lambda_6$,
\begin{equation}
w^3=\prod_{i=1}^6(z-\lambda_i) .\label{curvegen}
\end{equation}
Then the $\mathfrak{a}$-normalized Riemann period matrix is of the
form
\begin{align}
\hat\tau\sp{s}&=\rho^2\left(
H+(\rho\sp2-1)\frac{\boldsymbol{x}\boldsymbol{x}^T} {\boldsymbol{x}^T
H\boldsymbol{x}}   \right),\label{taumats}
\end{align}
where $H=\mathrm{diag}(1,1,1,-1)$. Then $\hat\tau\sp{s}$ is
positive definite if and only if
\begin{align} \bar{\boldsymbol{x}}^T H \boldsymbol{x} <0.
\label{condition1}
\end{align}
\end{proposition}
In fact the symmetry of (\ref{cyccurve}) means that $x_2=\rho
x_1$, $ x_3=\rho^2 x_1$, and only two periods need be computed.
Choosing the first sheet so that
$w=\sqrt[3]{(z^3-\alpha\sp3)(z^3+{1}/{\alpha\sp3})}$ is negative
and real on the real $z$-axis between the branch points
$(-1/\alpha,\alpha)$ these may be expressed in terms of the
integrals computed on the first sheet
\begin{align}
\mathcal{I}_1(\alpha)&=\int\limits_{0}^{\alpha}\frac{{d}z}{w}
=-\frac{2\pi\sqrt{3}\alpha}{9}
{_2F_1}\left(\frac13,\frac13;1;-\alpha^6\right),\\
\mathcal{J}_1(\alpha)&=\int\limits_{0}^{-1/\alpha}\frac{{d}z}{w} =
\frac{2\pi\sqrt{3}}{9\alpha} {_2F_1}\left(\frac13,\frac13;1;
-\alpha^{-6}\right).\label{integralsij}\end{align} Here
$_2F_1(a,b;c;z)$ is the standard Gauss hypergeometric function and
we may express the periods as
$x_{1}=-(2\mathcal{J}_1+\mathcal{I}_1)\rho -2\mathcal{I}_1
-\mathcal{J}_1$ and
$x_{4}=3(\mathcal{J}_1-\mathcal{I}_1)\rho+3\mathcal{J}_1$. Thus
the period matrix may be expressed via (\ref{taumats}) as rational
expressions in terms of $x_1$ and $x_4$, or equivalently in terms
of $\mathcal{I}_1$ and $\mathcal{J}_1$.

Now the Ercolani-Sinha constraints place constraints on the
periods. These were solved in \cite{bren06} to give
\begin{proposition}\label{propesnt}
To each pair of relatively prime integers $(m,n)=1$ for which
$$(m + n)(m -2n)<0$$
we obtain a solution\footnote{Note that $\boldsymbol{m}$ differs
by a sign from that of \cite{bren06} because we are working with
$\mathfrak{a}$-normalized quantities and their attendant
orientations (see the previous footnote).}
$\boldsymbol{n}=\begin{pmatrix} n& m-n&-m&2n-m
\end{pmatrix}$, $\boldsymbol{m}=\begin{pmatrix} -m& n&m-n&3n
\end{pmatrix}$ to the Ercolani-Sinha constraints for the curve
(\ref{cyccurve}) as follows. First we solve for $t$, where
\begin{equation}\label{esct}
\dfrac{2n-m}{m + n}=\frac{{_2F_1}(\frac{1}{3},
\frac{2}{3}; 1,t)}{{_2F_1}(\frac{1}{3}, \frac{2}{3}; 1,1-t)}.
\end{equation}
Then
\begin{equation}
b=\frac{1-2t}{\sqrt{t(1-t)}},\qquad t=
\frac{-b+\sqrt{b^2+4}}{2\sqrt{b^2+4}},
\end{equation}
and we obtain $\chi$ from
\begin{equation}\label{esevch}
\chi^{\frac{1}{3}} = -(n + m )\, \frac{2 \pi}{3
    \sqrt{3}}\ \frac{\alpha}{(1+\alpha\sp6)\sp\frac13}\ {_2F_1}(\frac{1}{3}, \frac{2}{3}; 1,
    t)
\end{equation}
with $\alpha\sp6=t/(1-t)$.
\end{proposition}
Remarkably one may solve the transcendental equation (\ref{esct})
using a theory developed to explain various formulae of Ramanujan
\cite{bbg95,bv98}. We shall not need any examples beyond those of
\cite{bren06}. At this stage one finds that the period matrix for
a curve (\ref{cyccurve}) satisfying \textbf{H1} and \textbf{H2}
may be expressed in terms in terms of the quantity
\begin{equation}\label{esR}\mathcal{R}=
\frac{\mathcal{I}_1(\alpha)}{\mathcal{J}_1(\alpha)}=
\frac{m-2n}{m+n}.\end{equation} We note the following symmetries
that preserve the constraints on $(m,n)$,
\begin{equation}\label{symnmR}
(m,n)\mapsto (-m,-n),\quad \mathcal{R}\mapsto \mathcal{R};\qquad
(m,n)\mapsto (n-m,n),\quad \mathcal{R}\mapsto \frac1{\mathcal{R}}.
\end{equation}
The remaining problem is to determine those allowed $(n,m)$ which
also satisfy \textbf{H3}. To make use of our alternative
characterisation (\ref{beh3}) we record
\begin{lemma}[\cite{bren06}] For the curve (\ref{cyccurve})
 $\widetilde{\boldsymbol{K}}=
\Theta_{{\rm singular}}$. Expressed as a characteristic,
$\widetilde{\boldsymbol{K}}
=\dfrac12\left[\begin{matrix}1&1&1&1\\1&1&1&1\end{matrix}\right]$.
\end{lemma}

\noindent{\bf Remark:} Because in the case under consideration
$\widetilde{\boldsymbol{K}}$ is an even half-period the function
the function  $\theta(\lambda\widehat{\boldsymbol{U}}\pm
\widetilde{\boldsymbol{K}},\hat\tau)$ vanishes to second order at
least at the points $\lambda=0$ and $\lambda=2$
\begin{align*}
\left.\theta(\lambda\widehat{\boldsymbol{U}}\pm
\widetilde{\boldsymbol{K}},\hat\tau)\right|_{\lambda\sim0}&=
\partial^2_{\widehat{\boldsymbol{U}}}
\theta(\widehat{\boldsymbol{K}};\widehat{\tau}) \lambda^2+O(\lambda^4),\\
\left.\theta(\lambda\widehat{\boldsymbol{U}}\pm
\widetilde{\boldsymbol{K}},\hat\tau)\right|_{\lambda\sim2}
&=\partial^2_{\widehat{\boldsymbol{U}}}
\theta(\widehat{\boldsymbol{K}};\widehat{\tau}) (\lambda-2)^2+O((\lambda-2)^4).
\end{align*}
We shall see that in fact it vanishes to order 4.

Finally we note some of the coverings associated with the curve
$\hat{\mathcal{C}}$.
\begin{lemma}\label{covering} The action of $\sigma$ on the curve (\ref{cyccurve})
yields the unramified covering
$\pi:\widehat{\mathcal{C}}\rightarrow\mathcal{C}:=\widehat{\mathcal{C}}/\texttt{C}_3$,
where $\mathcal{C}$ is the genus $2$  hyperelliptic curve
\begin{equation}
\mathcal{C}= \{ (\mu,\nu)| \nu^2=(\mu^3+b)^2+4   \},
\label{hcurve}
\end{equation}
with $\nu=z^3+1/z^3$ and $\mu=-w/z$. Further, $\mathcal{C}$
 two-sheetedly covers the two elliptic curves $\mathcal{E}_{\pm}$,
\begin{equation}
\mathcal{E}_{\pm}= \{ (z_{\pm},w_{\pm})|
w_{\pm}^2=z_{\pm}(1-z_{\pm})
(1-k_{\pm}^2z_{\pm})   \},\label{curvepm}\\
\end{equation}
where
\begin{align*}
  z_{\pm}&=\frac{K^2-L^2}{K^2-\rho L^2}\,
\frac{K-\mu}{\rho K-\mu}\,\frac{L^2-K\mu}{L^2-K\rho \mu},\\
w_{\pm}&=-\imath\sqrt{2+\rho}\sqrt{\frac{L\pm K}{L\mp K}}
\frac{K^2}{L} \frac{L^2-\rho K^2}{\rho L^2-K^2}\frac{\nu(L\mp
\mu)} {(\mu-\rho K)^2(L^2-\rho K \mu)^2
}.\label{cover2a}
\end{align*}
With $M={K}/{L}$, $K=(2\imath -b)^{\frac13}$ and
$L=(b^2+4)^{\frac16}$ the Jacobi moduli $k_{\pm}$ are given by
\begin{equation}\label{jacobimodpm}
k_{\pm}^2=-\frac{\rho(\rho M \pm 1)(\rho M \mp 1)^3  } {(M\pm
1)(M\mp 1 )^3}.
\end{equation}

\end{lemma}

\section{Fay-Accola factorization}
Thus far we have recalled earlier works: a countable putative
family of spectral curves for $SU(2)$ BPS monopoles has been
produced together with their period matrices and the vector $
\widetilde{\boldsymbol{K}}$ but it remains to discuss the Hitchin
constraint \textbf{H3}. The formulation of this constraint
(\ref{beh3}) is in terms of genus 4 theta functions and in this
section we wish to reduce this to questions of genus 2 theta
functions making use of a remarkable factorization theorem due to
Accola and Fay \cite{acc71, fay73} and also observed by Mumford.
Let $\pi:\hat{\mathcal{C}}\rightarrow\mathcal{C}$ be a cyclic
unramified covering. The map $\pi$ leads to a map $\pi\sp\ast:
\text{Jac}({\mathcal{C}})\rightarrow\text{Jac}(\hat{\mathcal{C}})$
which may be lifted to $\pi\sp\ast:{\mathbb C}\sp{g}\rightarrow
{\mathbb C}\sp{\hat g}$. When $\hat z=\pi\sp\ast z$ the theta
functions on $\hat{\mathcal{C}}$ and ${\mathcal{C}}$ are related
by this factorization theorem. We shall now describe this theorem
in the monopole setting.

Previously we have considered the symmetry $\mathrm{R}$ of the
spectral curve. Now we shall focus on the cyclic symmetry
$\sigma$, $\texttt{C}_3=<\sigma\,|\,\sigma^3=1>$ and the
unramified covering
$\pi:\widehat{\mathcal{C}}\rightarrow\mathcal{C}:=\widehat{\mathcal{C}}/\texttt{C}_3$
described in Lemma \ref{covering}. (The same considerations apply
to the curve $\eta^3+\alpha\eta\chi^2+\chi(\zeta^6+b \zeta^3-1)=0$
and more generally cyclically symmetric monopoles.) To implement
this theory we need a different choice of homology basis to that
described earlier which reflects this symmetry. We wish an
homology basis
$\{\hat{\mathfrak{a}}_0\sp{c},\ldots,\hat{\mathfrak{a}}_3\sp{c};
\hat{\mathfrak{b}}_0\sp{c},\ldots, \hat{\mathfrak{b}}_3\sp{c}\}$
on $\widehat{\mathcal{C}}$ and $\{\mathfrak{a}_0,\mathfrak{a}_1,
\mathfrak{b}_0,\mathfrak{b}_1\}$ on $\mathcal{C}$ satisfying (for
$i=1,2,3$)
\[ \sigma^k(\hat{\mathfrak{a}}_i\sp{c})=\hat{\mathfrak{a}}_{i+k}\sp{c},\;
\sigma^k(\hat{\mathfrak{b}}_i\sp{c})=\hat{\mathfrak{b}}_{i+k}\sp{c},\;
 \sigma^k(\hat{\mathfrak{a}}_0\sp{c})\sim\hat{\mathfrak{a}}_{0}\sp{c},\;
\sigma^k(\hat{\mathfrak{b}}_0\sp{c})=\hat{\mathfrak{b}}_{0}\sp{c},\quad
k=1,2,3, \] (that is $\sigma^k(\hat{\mathfrak{a}}_0\sp{c})$ is
homologous to $\hat{\mathfrak{a}}_0\sp{c}$) and such that
$$ \pi(\hat{\mathfrak{a}}_i\sp{c} )= {\mathfrak{a}}_1,\;
\pi(\hat{\mathfrak{b}}_i \sp{c})= {\mathfrak{b}}_1,\;
\pi(\hat{\mathfrak{a}}_0\sp{c} )= {\mathfrak{a}}_0,\;
\pi(\hat{\mathfrak{b}}_0\sp{c} )= 3{\mathfrak{b}}_0.$$ We may
construct such a basis as follows. We take
$\hat{\mathfrak{a}}_1\sp{c}=\hat{\mathfrak{a}}_1\sp{s}$,
$\hat{\mathfrak{b}}_1\sp{c}=\hat{\mathfrak{b}}_1\sp{s}=-\mathrm{R}\sp2
\hat{\mathfrak{a}}_1\sp{s}$ and
$\hat{\mathfrak{a}}_0\sp{c}=\hat{\mathfrak{a}}_4\sp{s}$ and extend
these by
$\hat{\mathfrak{a}}_{i+k}\sp{c}=\sigma_\ast\sp{k}(\hat{\mathfrak{a}}_i\sp{c})$
and
$\hat{\mathfrak{b}}_{i+k}\sp{c}=\sigma_\ast\sp{k}(\hat{\mathfrak{b}}_i\sp{c})$.
Thus
$\hat{\mathfrak{a}}_{i+k}\sp{s}=\mathrm{R}\sp{2(i+k-1)}\sigma_\ast\sp{k}(\hat{\mathfrak{a}}_i\sp{c})$,
and
$\hat{\mathfrak{b}}_{i+k}\sp{s}=\mathrm{R}\sp{2(i+k-1)}\mathrm\sigma_\ast\sp{k}(\hat{\mathfrak{b}}_i\sp{c})$.
At this stage we have defined the cyclic cycles
$\hat{\mathfrak{a}}_{i}\sp{c}$, $\hat{\mathfrak{b}}_{i}\sp{c}$
($i=1,2,3$) together with $\hat{\mathfrak{a}}_{0}\sp{c}$. We
complete the homology basis by seeking an invariant cycle
$\hat{\mathfrak{b}}_0\sp{c}$ and define the cycles on
$\mathcal{C}$ in terms of these. Such a basis is exhibited in
Figure 1(b). If we take as ordered bases
$\{\hat\gamma_i\sp{s}\}=\{\hat{\mathfrak{a}}_1\sp{s},\ldots,\hat{\mathfrak{a}}_4\sp{s};$
$ \hat{\mathfrak{b}}_1\sp{s},\ldots, \hat{\mathfrak{b}}_4\sp{s}\}$
and
$\{\hat\gamma_i\sp{c}\}=\{\hat{\mathfrak{a}}_1\sp{c},\ldots,\hat{\mathfrak{a}}_3\sp{c},
\hat{\mathfrak{a}}_0\sp{c}; \hat{\mathfrak{b}}_1\sp{c},\ldots,
\hat{\mathfrak{b}}_3\sp{c},\hat{\mathfrak{b}}_0\sp{c}\}$ then
$\hat\gamma_i\sp{c}=\mathfrak{S} \hat\gamma_i\sp{s}$ where
$\mathfrak{S}$ is the symplectic matrix
\begin{equation}\label{sympA}
\mathfrak{S}=\left(\begin {array}{cccccccc} 1&0&0&0&0&0&0&0\\\noalign{\medskip}0&
-1&0&0&0&1&0&0\\\noalign{\medskip}0&0&0&0&0&0&-1&0\\\noalign{\medskip}0
&0&0&1&0&0&0&0\\\noalign{\medskip}0&0&0&0&1&0&0&0\\\noalign{\medskip}0
&-1&0&0&0&0&0&0\\\noalign{\medskip}0&0&1&0&0&0&-1&0
\\\noalign{\medskip}0&0&0&-1&0&0&0&1\end {array}
\right):=\left( \begin{array}{cc} A&B\\C&D  \end{array}\right)
\in\mathrm{Sp}(8,\mathbb{Z}).
\end{equation}
For example
$$\hat{\mathfrak{a}}\sp{c}_2=\sigma\hat{\mathfrak{a}}\sp{c}_1=
\sigma\hat{\mathfrak{a}}\sp{s}_1=\mathcal{R}\hat{\mathfrak{a}}\sp{s}_2=
-\mathcal{R}\sp2\hat{\mathfrak{b}}\sp{s}_2=(1+\mathcal{R})\hat{\mathfrak{b}}\sp{s}_2
=-\hat{\mathfrak{a}}\sp{s}_2+\hat{\mathfrak{b}}\sp{s}_2.$$ Fay
works with the ordered basis
$\{\hat\gamma_i\sp{c}\}=\{\hat{\mathfrak{a}}_0\sp{c},
\hat{\mathfrak{a}}_1\sp{c},\ldots,\hat{\mathfrak{a}}_3\sp{c} ;
\hat{\mathfrak{b}}_0\sp{c},\hat{\mathfrak{b}}_1\sp{c},\ldots,
\hat{\mathfrak{b}}_3\sp{c}\}$, this reordering being achieved (on
both the $\mathfrak{a}$ and $\mathfrak{b}$-cycles) by
$$\mathrm{S}:= \left(
\begin {array}{cccc}
0&0&0&1\\
\noalign{\medskip}1&0&0&0\\
\noalign{\medskip}0&1&0&0\\
\noalign{\medskip}0&0&1&0\end {array}
 \right).
$$
We may again represent these cycles as integrals between branch
points:
\begin{align}\begin{split}
\hat{\mathfrak{a}}\sp{c}_1&=\gamma_1(\lambda_2,\lambda_1)+\gamma_2(\lambda_1,\lambda_2),
\qquad
\hat{\mathfrak{b}}\sp{c}_1=\gamma_1(\lambda_2,\lambda_1)+\gamma_3(\lambda_1,\lambda_2),\\
\hat{\mathfrak{a}}\sp{c}_2&=\gamma_2(\lambda_4,\lambda_3)+\gamma_3(\lambda_3,\lambda_4)
,\qquad
\hat{\mathfrak{b}}\sp{c}_2=\gamma_2(\lambda_4,\lambda_3)+\gamma_1(\lambda_3,\lambda_4),\\
\hat{\mathfrak{a}}\sp{c}_3&=\gamma_3(\lambda_6,\lambda_5)+\gamma_1(\lambda_5,\lambda_6)
,\qquad
\hat{\mathfrak{b}}\sp{c}_3=\gamma_3(\lambda_6,\lambda_5)
+\gamma_2(\lambda_5,\lambda_6),\\
\hat{\mathfrak{a}}\sp{c}_0&=\gamma_3(\lambda_2,\lambda_1)
+\gamma_2(\lambda_1,\lambda_5)+\gamma_3(\lambda_5,\lambda_6)
+\gamma_1(\lambda_6,\lambda_2),
\\
\hat{\mathfrak{b}}\sp{c}_0&
=\gamma_3(\lambda_1,\lambda_2)+\gamma_1(\lambda_2,\lambda_5)
+\gamma_2(\lambda_5,\lambda_6)
+\gamma_3(\lambda_6,\lambda_3)+\gamma_1(\lambda_3,\lambda_4)
+\gamma_2(\lambda_4,\lambda_1).
\end{split}\label{homology_c}
\end{align}
With the cyclic homology basis just described we have
\begin{theorem}[\bf Fay-Accola] \label{fayaccola}
With respect to the ordered canonical homology bases
$\{\hat\gamma_i\sp{c}\}=\{\hat{\mathfrak{a}}_0\sp{c},
\hat{\mathfrak{a}}_1\sp{c},\ldots,\hat{\mathfrak{a}}_3\sp{c} ;
\hat{\mathfrak{b}}_0\sp{c},\hat{\mathfrak{b}}_1\sp{c},\ldots,
\hat{\mathfrak{b}}_3\sp{c}\}$ and
$\{\mathfrak{a}_0,\mathfrak{a}_1, \mathfrak{b}_0,\mathfrak{b}_1\}$
specified above then the $\mathfrak{a}$-normalized Riemann period
matrices of $\hat{\mathcal{C}}$ and $\mathcal{C}$ take the
respective forms
\begin{equation}\hat{\tau}\sp{c}
=\left( \begin{array}{cccc} a&b&b&b\\
b&c&d&d\\
b&d&c&d\\
b&d&d&c
\end{array}\right),\qquad
\tau\sp{c}
=\left( \begin{array}{cc} \frac13 a&b\\
b&c+2d
\end{array}\right) .\end{equation}
Moreover for arbitrary $\boldsymbol{ z}=(z_0,z_1)\in \mathbb{C}^2$
then $\pi\sp\ast\boldsymbol{ z}=\boldsymbol{\hat
z}=(3\,z_0,z_1,z_1,z_1)$ and
\begin{equation}
\frac{\theta(3\,z_0,z_1,z_1,z_1;\hat{\tau}\sp{c})}
{\prod_{k=0}^{2}\theta\left[\begin{matrix}0&0
\\ \frac{k}{3}&0 \end{matrix}\right]\left(z_0,z_1;\tau\sp{c}
\right)}
=c_0(\widehat{\tau}\sp{c}
) \label{fafactora}
\end{equation}
is a non-zero modular constant
$c_0(\hat{\tau}\sp{c}
)
$ independent of
$\boldsymbol{ z}$.
\end{theorem}

In our setting we obtain
\begin{proposition}
The quantities $a,b,c,d$ appearing in the Fay-Accola theorem are
expressible in terms of the holomorphic integrals $x_1,x_4$ (with
$ \rho^3=1$) as
\begin{align}
a&=-\frac{6x_1^2-x_4^2+\rho(3x_1^2+x_4^2)}{3x_1^2-x_4^2},&
b&=\frac{(1+2\rho)x_1x_4}{3x_1^2-x_4^2},\\
c&=\frac{2x_1^2-x_4^2+\rho(x_1^2-x_4^2)}{3x_1^2-x_4^2},&
d&=-\frac{(1+2\rho)x_1^2}{3x_1^2-x_4^2}.
\label{abcd}
\end{align}
\end{proposition}
\begin{proof}
If we perform the symplectic transformation of the period matrix
(\ref{taumats}) with the symplectic transformation (\ref{sympA}),
$\hat\tau\sp{s}\rightarrow
(C+D\hat\tau\sp{s})(A+B\hat\tau\sp{s})^{-1}$, we obtain a period
matrix of the form
$$\left( \begin{array}{cccc}
c&d&d&b\\
d&c&d&b\\
d&d&c&b\\
b&b&b&a
\end{array}\right)=
\mathrm{S}\sp{-1}{\hat\tau}\sp{c} \mathrm{S},
$$
the final result coming after conjugation by $\mathrm{S}$ to
change the order of the homology basis to match that of Fay.
\end{proof}
Again we see that the period matrix just depends on the ratio of
$x_1/x_4$ or equivalently on $\mathcal{R}$. Now to make use of the
Fay-Accola theorem we must show that the vectors
$\widehat{\boldsymbol{U}}$ and $\widetilde{\boldsymbol K}$ may be
obtained by pullback from $\Jac(\mathcal{C})$. To this end we have

\begin{proposition} In the cyclic homology basis
$\{\hat{\mathfrak{a}}_0\sp{c},\ldots,\hat{\mathfrak{a}}_3\sp{c};
\hat{\mathfrak{b}}_0\sp{c},\ldots, \hat{\mathfrak{b}}_3\sp{c}\}$
the winding vector $\widehat{\boldsymbol{U}}$ and vector
$\widetilde{\boldsymbol{K}}$ take the form
\begin{align}
\widehat{\boldsymbol{U}}&=(\widehat{U}_0,\widehat{U}_1,\widehat{U}_1,
\widehat{U}_1),\quad
\widehat{U}_0=-\frac{C_0x_4}{3x_1^2-x_4^2},\quad \widehat{U}_1
=\frac{C_0 x_1}{3x_1^2-x_4^2},\label{wind}\\
\widetilde{\boldsymbol{K}}
&= \left( \frac12,\frac12,\frac12,\frac12\right)+
\left( \frac12,\frac12,\frac12,\frac12\right)\hat{\tau}\sp{c}
 =(\widetilde{K}_0,\widetilde{K}_1,
\widetilde{K}_1,\widetilde{K}_1),\label{ksymm}
\end{align}
where $C_0=-3(2n-m)$. The winding vector is a half-period and the
Ercolani-Sinha vector may be written $2\widehat{\boldsymbol{U}}=
\boldsymbol{\widehat n}+\boldsymbol{\widehat m}\,\hat{\tau}\sp{c}$
where
$$(\boldsymbol{\widehat n},\boldsymbol{\widehat m})=(5n-m,n,n,n,3n,-m,-m,-m).$$
\end{proposition}

\begin{proof} Using the explicit expression for the matrix  $\mathcal{A}$ from \cite{bren06}
one has the vector
$\widehat{\boldsymbol{U}}\sp{s}=\nu(1,0,0,0)\mathcal{A}^{-1} $ and
\begin{equation}
\widehat{\boldsymbol{U}}\sp{s}=-C_0\left( \frac{x_1}{x_4^2},\frac{\rho x_1}{x_4^2},
\frac{\rho^2 x_1}{x_4^2}, -\frac{1}{x_4}  \right).
\end{equation}
Then with the symplectic transformation (\ref{sympA})
$\widehat{\boldsymbol{U}}=\widehat{\boldsymbol{U}}\sp{s}
(A+B\hat\tau\sp{s})^{-1}$. The value of constant $C_0$ is found
from the condition $\bf{H2}$. Performing the symplectic
transformation on  the vector $(\boldsymbol{n},\boldsymbol{ m})$
given in Proposition 7 yields
$$(\boldsymbol{\widehat n}\mathrm{S},\boldsymbol{\widehat m}\mathrm{S})=(\boldsymbol{n},\boldsymbol{ m})
\mathfrak{S}\sp{-1}=(n,n,n,5n-m,-m,-m,-m,3n)
$$
and the result follows.

The only point to note is in the transformation of the vector of
Riemann constants. This has two parts: the vector $\boldsymbol K$
transforms as a vector, ${\boldsymbol K}\rightarrow{\boldsymbol
K}(A+B\hat\tau\sp{s})^{-1}$; but in transforming the argument of a
theta function function by a symplectic transformation the
characteristics of the theta function also transform (see Appendix
A), and a theta function with no characteristics may acquire
characteristics. When dealing with Riemann's theta function (with
vanishing characteristic) the acquired characteristics are
typically placed in the transformed vector of Riemann constants.
We do this here and find
$$\boldsymbol{\widetilde K}=(\frac12\ldots\frac12)\,\mathfrak{S}\sp{-1}
\begin{pmatrix}1\\ \widehat\tau\sp{c}
\end{pmatrix}+ \frac12\left((CD^T)_0,(AB^T)_0\right)
\begin{pmatrix}1\\ \widehat\tau\sp{c}\end{pmatrix}
$$
yielding the stated result.
\end{proof}
The form of the vectors $\widehat{\boldsymbol{U}}$ and
$\widetilde{\boldsymbol K}$ given in the theorem establishes
\begin{corollary}
$\widehat{\boldsymbol{U}}=\pi\sp\ast(\boldsymbol{U}\sp\ast)$ and
$\widetilde{\boldsymbol K}=\pi\sp\ast(\boldsymbol{K}\sp\ast)$
where $$ \boldsymbol{U}^{\ast}=\left(\frac13
\widehat{U}_0,\widehat{U}_1\right),\;
\boldsymbol{K}^{\ast}=\left(\frac13
\widehat{K}_0,\widehat{K}_1\right).$$
\end{corollary}
Therefore we may employ the Fay-Accola result. Introduce the three
functions
\begin{equation*}
f_{k}(\lambda)=\theta(\lambda\,\boldsymbol{U}^{\ast}+\boldsymbol{K}^{\ast}
+k\,\boldsymbol{l}^{\ast} \,|\,
\tau^c),\quad k=0,+1,-1, \quad\boldsymbol{l}^{\ast}=\left(\frac13,0\right).
\end{equation*}
Up to exponential factors these correspond to the three genus 2
theta functions with characteristics in the denominator of
(\ref{fafactora}). Making use of (\ref{beh3}, \ref{beh3b}) we
arrive at
\begin{theorem}\label{beH3} If $\lambda\in[0,2]$ then
\begin{equation}\label{beh3c}
H^0(\hat{\mathcal{C}},L^{\lambda}(n-2))\ne0\Longleftrightarrow
\theta(\lambda\widehat{\boldsymbol{U}}\pm
\widetilde{\boldsymbol{K}}\,|\,\hat\tau)=0\Longleftrightarrow
f_{k}(\lambda)=0
\end{equation} for at least one $k\in\{0,\pm1\}$.
\end{theorem}

At this stage we have reduced the question \textbf{H3} to
questions about various genus 2 theta functions and we shall look
at these in the next section.

\noindent{\bf{Remark:}} We note that the symplectic transformation
$$
\left( \begin {array}{cccccccc} 0&0&0&0&0&1&1&1\\
\noalign{\medskip}0&0&0&0&1&0&0&0\\
\noalign{\medskip}0&2&-1&-1&0&0&0&0\\
\noalign{\medskip}0&0&0&0&0&0&1&-1\\
\noalign{\medskip}0&-1&0&0&0&0&0&0\\
\noalign{\medskip}-1&0&0&0&0&0&0&0\\
\noalign{\medskip}0&0&0&0&0&0&0&-1\\
\noalign{\medskip}0&1&-1&0&0&0&0&0\end {array} \right)
$$
brings $\widehat\tau\sp{c}$ to the form
$$
\left( \begin {array}{cccc}
-\dfrac13\,{\dfrac {a}{-3\,{b}^{2}+ac+2\,ad}}&{\dfrac {b}{-3\,{b}^{2}+ac+2\,ad}}&-\dfrac13&0\\
\noalign{\medskip}{\dfrac {b}{
-3\,{b}^{2}+ac+2\,ad}}&-{\dfrac {c+2\,d}{-3\,{b}^{2}+ac+2\,ad}}&0&0\\
\noalign{\medskip}-\dfrac13&0&\dfrac{c-d}6&\dfrac12\\
\noalign{\medskip}0&0&\dfrac12&-\dfrac12\, \dfrac1{ c-d }
\end {array} \right).
$$
One may identify the top $2\times2$ block as conjugate to
$-\tau\sp{c\,-1}/3$. Using this one can use Weierstrass reduction
to rewrite the genus 4 theta functions as in \cite{bren06}.

\section{The Humbert variety}

At this stage by using the Fay-Accola theorem we have reduced the
question of the vanishing of the genus 4 $\theta$-function
$\theta(\lambda\,\widehat{\boldsymbol{U}}+
\widetilde{\boldsymbol{K}};\widehat{\tau}\sp{c} )$ to that of the
vanishing of the genus 2 $\theta$-functions $f_k(\lambda)$,
$k\in\{0,\pm1\}$. We can in fact do better. In \cite{bren06} it
was observed that $\mathcal{C}$ covered two elliptic curves and we
shall now exploit this geometry making use of ideas of Humbert
expounded in Krazer \cite{kr03} that we now recall.

\begin{definition}
The period matrix $\tau$ of a genus two algebraic curve
$\mathcal{C}$ belongs to the Humbert variety
$\mathcal{H}_{\Delta}$ associated with the symplectic invariant
$\Delta$ if there exist integer $q_i\in\mathbb{Z}$ satisfying
\begin{equation} q_1+q_2\tau_{11}+q_3\tau_{12}+q_4\tau_{22}+
q_5 (\tau_{12}^2-\tau_{11}\tau_{22})=0 \label{humbert}
\end{equation} and
\begin{equation}
\label{hopf} q_3^2-4(q_1q_5+q_2q_4)=\Delta.
\end{equation}
The curve $\mathcal{C}$ covers elliptic curves $\mathcal{E}_{\pm}$
if and only if $\Delta$  is a perfect square, $\Delta=h^2\geq 1$,
$h\in\mathbb{N}$. Then the integer $h$ is the degree of the cover.
\end{definition}

\begin{theorem}[\bf Bierman-Humbert] Let $\tau\in \mathcal{H}_{\Delta}$ and $\Delta=h^2$.
Then there exists a symplectic transformation $\mathfrak{S}\in
\mathrm{Sp}(4,\mathbb{Z})$, such that
\begin{equation} \mathfrak{S}\circ \tau =\widetilde{\tau}=\left( \begin{array}{cc} \widetilde{\tau}_{11}&\frac{1}{h}\\
\frac{1}{h}&\widetilde{\tau}_{22}
 \end{array}   \right) \label{beir-humb} \end{equation}
The transformation $\mathfrak{S}$ is given constructively and may
be realized in a finite number of steps.
\end{theorem}
A modern proof of the theorem is given \cite{murabayashi94}
revising that of Krazer \cite{kr03}.

When a $2\times2$ period matrix $\widetilde{\tau}$ has the
structure (\ref{beir-humb}) we may decompose the associated
$\theta$-function as
\begin{align}
\theta(z_1,z_2\,|\,\widetilde{\tau})=
\sum_{k=0}^{h-1}
\vartheta_3\left(z_1+\frac{k}{h}\,|\,\widetilde{\tau}_{1,1}\right)\theta
\left[\begin{array}{c}\frac{k}{h}\\0  \end{array}\right]
\left( hz_2 \,|\,h^2\widetilde{\tau}_{2,2}\right), \quad h^2=\Delta.\label{decomposition}
\end{align}
Here and below $\vartheta_k(z|\tau)$, $k=1,2,3,4$ denote the
Jacobi theta functions \cite{ba55}.

Our case is relevant to the Humbert variety $\mathcal{H}_{4}$ that
has received most study. The following is true.
\begin{proposition} Let $\tau\sp{c}$ be the period matrix of the curve $\mathcal{C}$.
Then $\tau\sp{c}\in \mathcal{H}_{4}$ with
\begin{equation}
\widetilde{\tau}_{11}=
\frac12\,{\frac { \left( \rho-1 \right)
 \left( -3\,{x}_{{1}}+2\,{x}_{{4}}+\rho\,{x}_{{4}}
 \right) }{3\,{x}_{{1}}-{x}_{{4}}+\rho\,{x}_{{4}}}}
, \quad
\widetilde{\tau}_{22}=
\frac16\,{\frac { \left( 2+\rho \right)  \left( -3
\,{x}_{{1}}-{x}_{{4}}+\rho\,{x}_{{4}} \right) }{3\,{
x}_{{1}}+2\,{x}_{{4}}+\rho\,{x}_{{4}}}}
.\label{tau1122a}
\end{equation}
\end{proposition}
\begin{proof} Substituting the expressions we have for $a$, $b$, $c$ and
$d$ in terms of $x_1$ and $x_4$ in the matrix equality
\[ \tau\sp{c}=\left(\begin{array}{cc} \tau_{11}&\tau_{12}\\
\tau_{12}&\tau_{22}
 \end{array} \right)=\left(\begin{array}{cc}\frac13 a&b\\b&c+2d
 \end{array} \right) \]
we may eliminate $x_1$ and $x_4$ to obtain the two relations:
\begin{align}
0&=-2-\tau_{22}+3\tau_{11}\label{relation1a},\\
0&=1-3\tau_{11}-\tau_{12}^2
\emph{}+3\tau_{11}^2.\label{relation2a}
\end{align}
Using the first of these we may write $3\tau_{11}^2=
\tau_{11}(\tau_{22}+2)$ which leads to the second taking the form
\[1-\tau_{11}+\tau_{11}
\tau_{22}-{\tau_{12}}^2=0.\] This is (\ref{humbert}) with $q_1 =
1$, $q_2 = -1$, $q_3 = 0$, $q_4 =0$, $q_5 = -1$ and the value of
the invariant $\Delta=4$. (We remark that other possibilities may
arise in the elimination process but we present only one resulting
in $\Delta=4$.) Standard procedure \cite{kr03}, \cite{bbeim94}
yields the symplectic transformation
\begin{equation}
\mathfrak{S}=
\left(\begin{array}{cccc}
 0&1&1&0\\
 \noalign{\medskip}1&1&0&1\\
\noalign{\medskip}0&1&0&1\\
\noalign{\medskip}0&0&1&0\end {array}
\right)=\left( \begin{array}{cc}
\alpha&\beta\\ \gamma&\delta\end{array} \right)
\in \mathrm{Sp}(4,\mathbb{Z})\label{transf2a}
\end{equation}
which reduces $\tau\sp{c}$ to the form (\ref{beir-humb}) with
$h=2$ and the stated identifications (\ref{tau1122a}).
\end{proof}

This proposition enables us to write the genus two theta functions
and so the functions $f_k$ in terms of Jacobi $\theta$-functions,
\begin{align}\label{tfa}
\theta(z_1,z_2\,|\,\widetilde{\tau})=
\vartheta_3\left(z_1\left.\right|\widetilde{\tau}_{1,1}\right)\vartheta_3
\left( 2z_2\,|\,4\widetilde{\tau}_{2,2}\right)+
\vartheta_3\left(z_1+1/2\left.\right|\widetilde{\tau}_{1,1}\right)\vartheta_2
\left( 2z_2\,|\,4\widetilde{\tau}_{2,2}\right).
\end{align}
We will need to transform the argument ${\boldsymbol
z}=\lambda\,\boldsymbol{U}^{\ast}+\boldsymbol{K}^{\ast}
+k\,\boldsymbol{l}^{\ast}$ using the transformation
(\ref{transf2a}) but before doing so it is helpful to consider the
moduli $\widetilde{\tau}_{11}$ and $\widetilde{\tau}_{22}$ and an
additional link between them. As explained earlier, the periods
$x_{1,4}$ are expressible in terms of the integrals
$\mathcal{I}_1(\alpha)$ and $\mathcal{J}_1(\alpha)$ whose ratio
(\ref{esR}) is constrained to be $\mathcal{R}$. Only the ratios of
$x_{1,4}$ appear in (\ref{tau1122a}) and we may replace these by
$\mathcal{R}$,
$$\widetilde{\tau}_{11}=\frac12\,{\frac
{2+4\,\rho+3\,\mathcal{R}}{\mathcal{R}+1+2\,\rho}}
=\frac{2i\sqrt{3}+3\,\mathcal{R}}{2(\mathcal{R}+i\sqrt{3})}
,\qquad
\widetilde{\tau}_{22}=-{\frac {1+2\,\rho+3\,\mathcal{R}}{6\,\mathcal{R}}}=
-\frac{i\sqrt{3}+3\,\mathcal{R}}{6\,\mathcal{R}}.
$$
It is convenient to introduce the purely imaginary quantity (with
positive imaginary part)
\begin{equation}
\mathcal{T}=-\frac{2\imath\sqrt{3}}{\mathcal{R}}=2\imath\sqrt{3}\,
\frac{n+m}{2n-m}.\label{finmoda}
\end{equation}
In terms of this we have
\begin{equation}\label{modTa}
\widetilde{\tau}_{11}=1- \frac1{\mathcal{T} -2},\qquad
\widetilde{\tau}_{22}=\frac{\mathcal{T}}{12}-\frac12
\end{equation}
and the transformed arguments take the form
\begin{align*}
\boldsymbol{U}'&= {\boldsymbol{U}}^{\ast}
(\alpha+\beta\tau)^{-1}=\left(
\frac { \left( -1+i\sqrt {3} \right) C_0\,\mathcal{T} }{36(\mathcal{T}-2)},
-\frac{\left( 3+i\sqrt {3} \right) C_0\,\mathcal{T} }{216}
\right),\\
\boldsymbol{l}'&=\boldsymbol{l}^{\ast}(\alpha+\beta\tau)^{-1}=\left(
-\frac { \left( \mathcal{T}-3 \right)}{3(\mathcal{T}-2)}, \frac16
\right), \\
\boldsymbol{K}'&={\boldsymbol{K}}^{\ast}(\alpha+\beta\tau)^{-1}+
\frac12\left((\gamma\delta^T)_0,(\alpha\beta^T)_0\right)
\begin{pmatrix}1\\ \tau_{\mathfrak{a}}\end{pmatrix}
=\left(\frac43-\frac1{3(\mathcal{T}-2)},\frac{\mathcal{T}}{12}-\frac16\right).
\end{align*}
Using (\ref{modTa}, \ref{tfa}) and various substitutions the
following proposition is established in Appendix B.
\begin{proposition}\label{humred} For each pair of relatively prime integers
$(m,n)=1$ for which $(2n-m)(n+m)>0$ let $\widehat{\boldsymbol{U}}$
be the Ercolani-Sinha vector and $\widehat{\tau}$ the period
matrix of the genus 4 curve described above. Then the function
$\theta(\lambda\,\widehat{\boldsymbol{U}}+\widehat{\boldsymbol{K}}\,|\,
\widehat{\tau} )$ vanishes for $\lambda\in[0,2]$ if and only if at
least one of the three functions (with $k\in\mathbb{Z}$)
\begin{align}
h_k(y):=\frac{\vartheta_3}{\vartheta_2}
\left(i\sqrt{3}\,y+\frac{k\,\mathcal{T}}{3}\,\Big|\,\mathcal{T}\right)
+(-1)^{k}\,
\frac{\vartheta_2}{\vartheta_3}\left( y+ \frac{k}{3}
\,|\,\frac{\mathcal{T}}{3}\right),\quad k=-1,0,1\mod 3,\label{vanishing11a}
\end{align}
also vanishes. Here $y:=y(\lambda)=\lambda\,(n+m)\rho/3$,
$\mathcal{T}=2i\sqrt{3}(n+m)/(2n-m)$ and
$\frac{\vartheta_3}{\vartheta_2} \left(z|\mathcal{T}\right)$ is
shorthand for $\frac{\vartheta_3\left(z|\mathcal{T}\right)}{
\vartheta_3\left(z|\mathcal{T}\right)}$. Further the functions
$h_k$ satisfy \begin{align}\label{hper} h_{k+3}(y)&=h_k(y), \qquad
h_{k}\left(y+\frac{2(n+m)}3\right)=h_{k-[n+m]}(y),\nonumber \\
h_{k}\left(y+\frac{2(n+m)}3\rho\right) &=\begin{cases}
(-1)\sp{n+m}\,h_{k-[n+m]}(y)&\text{if }m\text{ even},\\
(-1)\sp{k}h_{k-[n+m]}(y+\mathcal{T}/2) &\text{if }m\text{ odd},
\end{cases}\\
h_k(y(\lambda+2))&=0 \Longleftrightarrow
h_{k-[n+m]}(y(\lambda))=0.\nonumber
\end{align}
\end{proposition}
Therefore $h_k$ are elliptic functions with periods $2(n+m)$ and
$4(n+m)\rho$. We also note that the zero divisors of $h_k(y)$ and
$h_k(y+\mathcal{T}/2)$ are the same.

Thus we have reduced the question of \textbf{H3} to that of the
zeros of the elliptic functions $h_k$. This theta function
question is much simpler than the corresponding (much greater
degree) theta function expressions of \cite{bren06} which arose
making use of Weierstrass-Poincar\'e reduction. Exploiting the
geometry has greatly simplified the problem. We shall turn to the
theta function question in the next section.

\noindent{\bf{Remark:}} We have an action of $\Gamma(2)\times
\Gamma(2)$ on period matrices of the form
$\begin{pmatrix}\lambda_1&1/2\\ 1/2&\lambda_2\end{pmatrix}$ which
may be associated to any genus 2 curve with extra involution
distinct from the hyperelliptic involution. Here
$\Gamma(2)=\left\{\begin{pmatrix}a&b\\c&d\end{pmatrix}\in
PSL(2,\mathbb{Z})\,\Big|\, a\equiv d\equiv1\pmod2,\ b\equiv
c\equiv0\pmod2 \right\}$ has generators $\tau\mapsto \tau+2$ and
$\tau\mapsto \frac{\tau}{1-2\tau}$ and we have the exact sequence
$$1\rightarrow\Gamma(2)\rightarrow PSL(2,\mathbb{Z})\rightarrow
S_3\rightarrow1.$$ To see this we observe that with the action
$\begin{pmatrix}A&B\\C&D\end{pmatrix}$:
$\tau\mapsto(C+D\,\tau)(A+B\,\tau)\sp{-1}$ we have
\begin{align*}
s&&\begin{pmatrix}
1&2&-4&0\\
0&1&0&0\\
0&0&1&0\\
0&1&-2&1
\end{pmatrix}&&\begin{pmatrix}\lambda_1&1/2\\ 1/2&\lambda_2\end{pmatrix}
\mapsto
\begin{pmatrix}\frac{\lambda_1}{1-4\lambda_1}&1/2\\ 1/2&\lambda_2\end{pmatrix}\\
t&&\begin{pmatrix}
1&0&0&0\\
0&1&0&0\\
1&0&1&0\\
0&0&0&1
\end{pmatrix}&&\begin{pmatrix}\lambda_1&1/2\\ 1/2&\lambda_2\end{pmatrix}
\mapsto
\begin{pmatrix}{\lambda_1}+1&1/2\\ 1/2&\lambda_2\end{pmatrix}\\
&&\begin{pmatrix}
0&1&0&0\\
1&0&0&0\\
0&0&0&1\\
0&0&1&0
\end{pmatrix}&&\begin{pmatrix}\lambda_1&1/2\\ 1/2&\lambda_2\end{pmatrix}
\mapsto\begin{pmatrix}\lambda_2&1/2\\ 1/2&\lambda_1\end{pmatrix}
\end{align*}
If we set $\mu=2\lambda_1$ then $s$ and $t$ give the actions
$\mu\mapsto \mu+2$, $\mu\mapsto \frac{\mu}{1-2\mu}$ and so
generate $\Gamma(2)$; conjugation by the final matrix then extends
this to $\Gamma(2)\times \Gamma(2)$. Then with
\begin{equation}\label{bolzaform}
\widetilde{\tau}_{11}'=s^2t^{-1}\left(\widetilde{\tau}_{11}\right)
=- \frac1{\mathcal{T} +6},
\qquad
\widetilde{\tau}_{22}'=t\left(\widetilde{\tau}_{22}\right)=\frac{\mathcal{T} +6}{12}
\end{equation}
and we have $12\widetilde{\tau}_{11}'\widetilde{\tau}_{22}'+1=0$,
the relation Bolza's claimed for period matrices for such curves
\cite{b88}.

We may now give an alternate characterization of those curves
(\ref{bren03}) satisfying Hitchin's conditions $\textbf{H1}$ and
$\textbf{H2}$.
\begin{proposition}\label{formbmn}
The family of curves $\eta^3+\chi(\zeta^6+b\zeta^3-1)=0$ satisfy
the constraints $\mathbf{H1}$ and $\mathbf{H2}$ when
\begin{equation}
b(m,n)=-\frac{\sqrt{3}(p(m,n)^6-45p(m,n)^4+135p(m,n)^2-27) }
{9p(m,n)(p(m,n)^4-10p(m,n)^2+9)}\label{bformula}
\end{equation}
and $\chi=\chi(m,n)$ may be expressed in terms of $m$, $n$ and
$b(m,n)$ by Proposition \ref{propesnt}. Here $m$ and $n$ are
relatively prime integers $(m,n)=1$ for which $(m+n)(m-2n)<0$ and
\begin{equation} p(m,n)=\frac{ 3\vartheta_3^2
\left(0\vert \frac{\mathcal{T}(m,n)}{2}\right) }
{\vartheta_3^2\left(0\vert \frac{\mathcal{T}(m,n)}{6}\right)
},\qquad
\mathcal{T}(m,n)=2\imath\sqrt{3}\frac{n+m}{2n-m}.\end{equation}
\end{proposition}
Indeed we may relate the elliptic curves $\mathcal{E}_\pm$ of
Lemma \ref{covering} with the period matrix (\ref{tau1122a}) or
the symplectically equivalent (\ref{bolzaform}) via
\begin{corollary}\label{modulikpm}
The genus two curve $\mathcal{C}$ 
two-sheetedly covers the elliptic curves $\mathcal{E}_{\pm}$ whose
Jacobian moduli may be written
\[ k_+^2=\frac{\vartheta_2^4\left(0\vert \frac{\mathcal{T}}{6}\right)}
{\vartheta_3^4\left(0\vert \frac{\mathcal{T}}{6}\right)},\quad
k_-^2=\frac{\vartheta_2^4\left(0\vert \frac{\mathcal{T}}{2}\right)}
{\vartheta_3^4\left(0\vert \frac{\mathcal{T}}{2}\right)} \]
\end{corollary}
The proof of both the proposition and corollary is presented in
Appendix C.

\section{The Theta function Question}
The final step in establishing the existence of monopoles with
spectral curve (\ref{bren03}) is then to understand the vanishing
properties of the function
\begin{equation}\label{defH}
H(y)=h_{-1}(y) h_0(y) h_1(y)
\end{equation}
with the definitions introduced in Proposition \ref{humred}.
$H(y)$ is also an elliptic function with periods $2(n+m)/3$ and
$4(n+m)\rho/3$. Given the periodicity in $k$ of $h_k$ proven in
this proposition we have that
\begin{lemma} $H(y(\lambda))=0\Leftrightarrow H(y(\lambda+2))=0$ and these
functions have the same vanishing properties.
\end{lemma}
Numerical calculations in \cite{bren07} suggested the conjecture
\begin{conjecture}\label{conjecture}
For each pair of relatively prime integers
$(m,n)=1$ for which $(2n-m)(n+m)>0$ let
$y=y(\lambda)=\lambda(n+m)\rho/3$ and
$\mathcal{T}=2i\sqrt{3}(n+m)/(2n-m)$. Then $H(y)$ vanishes
$2(|n|-1)$ times on the interval $\lambda\in(0,2)$.
\end{conjecture}

To prove the uniqueness of the tetrahedral monopole within the
class of symmetric monopole curves it will suffice to show only
$(m,n)=(1,1)$ and $(0,1)$ have no zeros within the range. At
present we don't know how to prove the more general conjecture. We
expect vanishing at $\lambda=0$ and $2$. This follows here due to
\begin{lemma}\label{fundv}
We have the following identities for all $\tau$ in the upper
half-plane:
\begin{align}
\frac{\vartheta_3\left(\frac{\tau}{3}\left.\right|\tau \right)}
{\vartheta_2\left(\frac{\tau}{3}\left.\right|\tau \right)}&=
\frac{\vartheta_2\left(\frac{1}{3}\left.\right|\frac{\tau}{3} \right)}
{\vartheta_3\left(\frac{1}{3}\left.\right|\frac{\tau}{3} \right)}\label{relation}\\
\vartheta_4^2(0|\tau)i\sqrt{3}&\,
\frac{ \vartheta_1\left.\left(\frac{\tau}{3}\right|\tau \right)
\vartheta_4\left.\left(\frac{\tau}{3}\right|\tau \right)  }
{\vartheta_2^2\left.\left(\frac{\tau}{3}\right|\tau \right)}+
\vartheta_4^2\left(0\left|\frac{\tau}{3}\right)\right.
\frac{ \vartheta_1\left.\left(\frac{1}{3}\right|\frac{\tau}{3} \right)
\vartheta_4\left.\left(\frac{1}{3}\right|\frac{\tau}{3} \right)  }
{\vartheta_3^2\left.\left(\frac{1}{3}\right|\frac{\tau}{3} \right)}=0\label{derrel}
\end{align}
Similar identities may be obtained by cyclic interchange of the
$\theta$-subscripts $i,j,k\in \{2,3,4\}$.
\end{lemma}
As a consequence we obtain
\begin{align*}
\frac{\vartheta_3\left(\frac{\tau}{3}\left.\right|\tau \right)}
{\vartheta_2\left(\frac{\tau}{3}\left.\right|\tau \right)}=
\frac{\vartheta_3\left(\pm\frac{\tau}{3}\left.\right|\tau \right)}
{\vartheta_2\left(\pm\frac{\tau}{3}\left.\right|\tau \right)}=
\frac{\vartheta_3\left(\frac{2\tau}{3}\left.\right|\tau \right)}
{\vartheta_2\left(\frac{2\tau}{3}\left.\right|\tau \right)}&=
\frac{\vartheta_2\left(\frac{1}{3}\left.\right|\frac{\tau}{3} \right)}
{\vartheta_3\left(\frac{1}{3}\left.\right|\frac{\tau}{3} \right)}=-
\frac{\vartheta_2\left(\frac{2}{3}\left.\right|\frac{\tau}{3} \right)}
{\vartheta_3\left(\frac{2}{3}\left.\right|\frac{\tau}{3} \right)}.
\end{align*}
 Although we have not seen these
identities in the standard texts known to us these identities may
be established by standard techniques. We then have,

\begin{lemma}\label{vanishing02}
At $\lambda=0$ we have
\begin{equation}
h_{\pm1}(0)=0, \; h_0(0)\neq 0,\label{vanishing111}
\end{equation}
and each of the functions, $h_{\pm1}(y(\lambda))$ vanish to second
order in $\lambda$.
\end{lemma}

\begin{proof} At the point $y=0$ we have that $h_{\pm1}(0)=0$ on
account of (\ref{relation}). Further the derivatives
\begin{align*}
\frac{d}{dz}\left(\frac{\vartheta_3}{\vartheta_2}\left(z\,\Big|\,\mathcal{T}\right)\right)&=\pi
\vartheta_4^2(0\vert\mathcal{T})\,
\frac{\vartheta_1}{\vartheta_2}\left(z\,\Big|\,\mathcal{T}\right)
\frac{\vartheta_4}{\vartheta_2}\left(z\,\Big|\,\mathcal{T}\right),\\
\frac{d}{dz}\left(\frac{\vartheta_2}{\vartheta_3}\left(z\,\Big|\,\mathcal{T}\right)\right)&=-\pi
\vartheta_4^2(0\vert\mathcal{T})\,
\frac{\vartheta_1}{\vartheta_3}\left(z\,\Big|\,\mathcal{T}\right)
\frac{\vartheta_4}{\vartheta_3}\left(z\,\Big|\,\mathcal{T}\right).
\end{align*}
show that $h_{\pm1}'(0)=0$ also vanishes on account of
(\ref{derrel}) and so both vanish to second order here. Standard
$\theta$-function expansions show that $h_0(0)$ is nonvanishing.

\end{proof}

A consequence of this lemma is that $H(y(\lambda))$ vanishes to
fourth order in $\lambda$ at $\lambda=0$ and $2$; this is
equivalent to the higher order vanishing of
$\theta(\lambda\widehat{\boldsymbol{U}}\pm
\widetilde{\boldsymbol{K}},\hat\tau)$ remarked upon earlier. We
may now establish the theorem stated in the introduction.

\begin{proof}[Proof of Theorem 1]
The dependence (\ref{finmoda}) of the modulus
$\mathcal{T}=\mathcal{T}(\mathcal{R})$ on the ratio $\mathcal{R}$
means that the vanishing $h_k(y)=0$ and $H(y)=0$ define
(respectively) implicit functions $y=X_k(\mathcal{R})$ and
$y=X(\mathcal{R})$ of the real variable $\mathcal{R}$. Although
our problem has $\mathcal{R}\in\mathbb{Q}$ we may extend its
domain to the whole real half-line, $|\mathcal{R}|\in\mathbb{R}^+$
(recall our conventions are such that $\mathcal{R}<0$). The
functions $X_\ast(\mathcal{R})$ are clearly multi-valued
reflecting the periodicities of $h_k(y)$ and $H(y)$. We may
determine many points on $X(\mathcal{R})$ using the fundamental
Lemma \ref{fundv}. For example, consider $y=2\rho/3$ and solutions
to $h_{-1}(2\rho/3)=0$. Substitution and some simplification leads
to solving
\begin{equation}\label{mudots}\frac{\vartheta_3}{\vartheta_2}
\left(\frac{|\mathcal{R}|+2}6\,\mathcal{T}\,\Big|\,\mathcal{T}\right)
=\frac{\vartheta_2}{\vartheta_3}\left(
\frac{|\mathcal{R}|}6\,\mathcal{T}+ \frac{1}{3}
\,|\,\frac{\mathcal{T}}{3}\right).
\end{equation}
Using Lemma \ref{fundv} we find solutions when
\begin{itemize}
\item $|\mathcal{R}|$ is even, giving $|\mathcal{R}|=6k$ or
    $6k+2$ ie $2,6,8,12,14,\ldots$
\item $|\mathcal{R}|$ is odd, giving
    $|\mathcal{R}|=3,5,9,11,15,17,\ldots$.
\end{itemize}
Similar arguments give solutions $$\begin{array}{lcl}
y=2\rho/3&h_{-1}&|\mathcal{R}|=2,3,5,6,8,9,11,12,\ldots\\
y=2\rho/3&h_{0}&|\mathcal{R}|=1,2,4,5,7,8,10,11,\ldots\\
y=4\rho/3&h_{0}&|\mathcal{R}|=1,2,4,5,7,8,10,11\ldots\\
y=4\rho/3&h_{0}&|\mathcal{R}|=3k+1/2,\\
y=4\rho/3&h_{1}&|\mathcal{R}|=2,3,5,6,8,9,11,12,\ldots\\
y=4\rho/3&h_{1}&|\mathcal{R}|=3k+1/2,\ldots\\
y=2\rho&h_{1}&|\mathcal{R}|=1,2,3,4,5,6,7,8,9,\ldots\\
\end{array}$$
and so on. At several of these points these points the tangent to
$X_\ast(\mathcal{R})$ becomes vertical; these may be obtained by
solving the analogous formulae for the tangent. A graph of some of
the components of $X(\mathcal{R})$ is given in Figure
\ref{figXR3}.

\begin{figure}[scale=1.25]
\caption{The plot shows (some) branches of the
multi-valued function $y=X(\mathcal{R})$ given implicitly by the equation $H(y)=0$.
Circles on the plot shows points at which the tangent lines are vertical.
The bold lines correspond to $|\mathcal{R}|=2$ and $1/2$. The different colours
correspond to branches of $y=X_k(\mathcal{R})$ (blue=$X_1$, green=$X_0$, red=$X_1$).}
\label{figXR3}
\vskip1cm
  \begin{center}
  \setlength{\unitlength}{1cm}
\begin{picture}(11,8)
  \put(-1,4){$y/\rho$}
  \put(6,-1){$|\mathcal{R}|$}
  \put(0,0){\includegraphics[width=11cm,height=8cm]{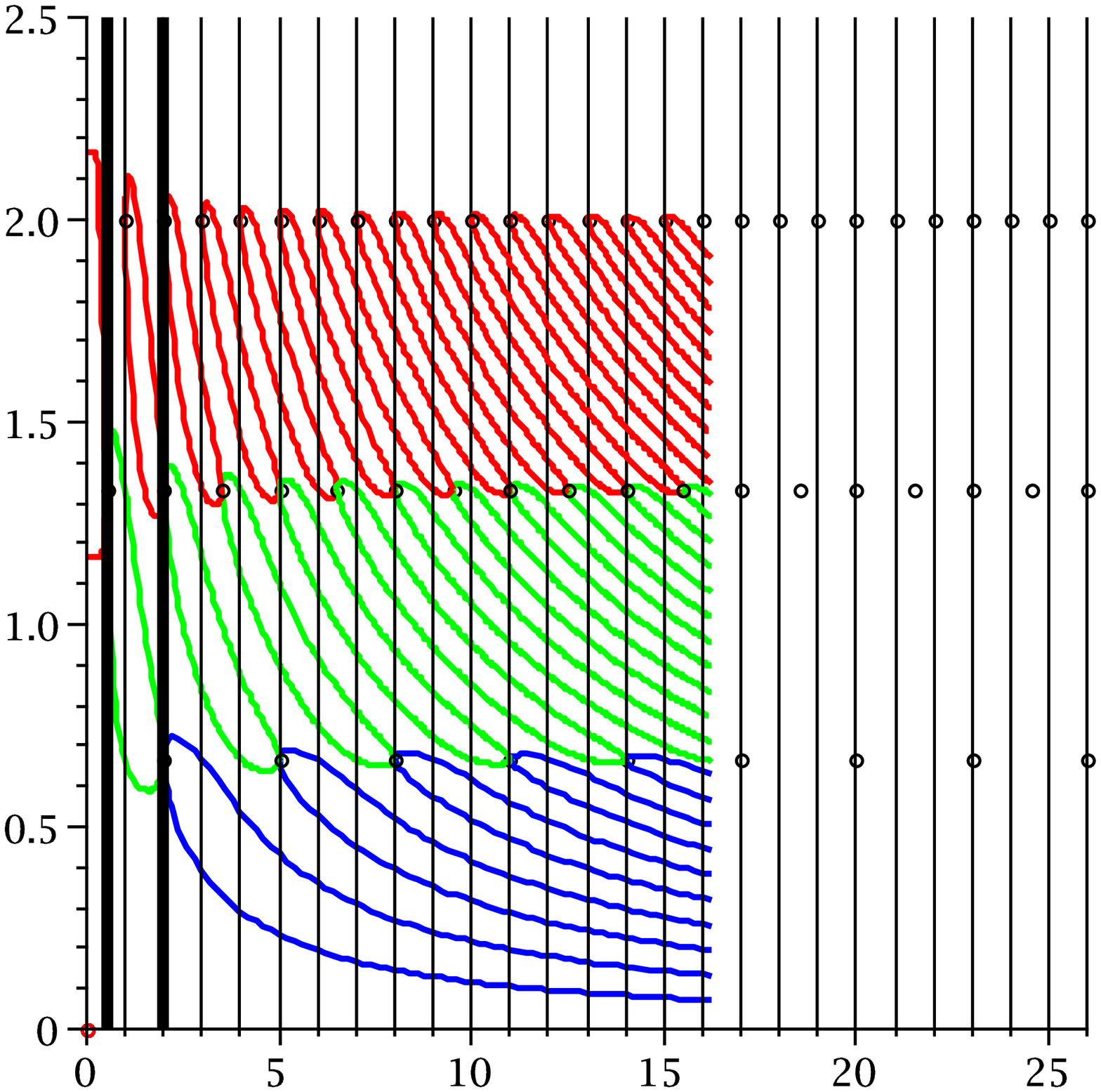}}
\end{picture}
\end{center}
\end{figure}

Now using the symmetry (\ref{symnmR}) we may assume that $n+m\ge1$
and so for $\lambda\in(0,2)$ then
$y/\rho\in(0,2(n+m)/3)\supseteq(0,2/3)$. Now we see that
$X(\mathcal{R})$ always has a zero in $(0,2/3)$ for all
$|\mathcal{R}|=(2n-m)/(n+m)\in(1,2)\cup(2,\infty)$ and so these
values cannot yield a monopole. Similarly if $n+m\ge2$ there is
always a zero of $X(\mathcal{R})$ in $(0,4/3)$ for any
$|\mathcal{R}|\ne1/2$. Thus we must have either $n+m=1$ and
$|\mathcal{R}|\in(0,1]\cup\{2\}$ or $n+m=2$ and
$|\mathcal{R}|=1/2$. The only solutions to these constraints are
$(m,n)=(0,1)$ with $|\mathcal{R}|=2$ and $(m,n)=(1,1)$ with
$|\mathcal{R}|=1/2$. For all other $(m,n)$ there are solutions to
$H(y)=0$ for $\lambda\in(0,2)$ and thus by Proposition
\ref{humred} they do not yield monopoles. Now the two cases
$(m,n)=(0,1)$, $(1,1)$ were shown to yield the tetrahedrally
symmetric monopoles in \cite{bren06}. Thus we have established
Theorem \ref{mainthm}.
\end{proof}

\noindent{\bf{Remark:}} We numerically observe that
$h_k(y)=0\Longleftrightarrow h_k(\rho y)=0 \Longleftrightarrow
h_k(\rho^2 y)=0$. The modulus of the elliptic functions $h_k$ is
$2\rho$ so this is not simply complex multiplication. This
observation remains unexplained.

\section{Discussion}
This paper has been devoted to the study of certain charge three
(centred) $SU(2)$ BPS monopoles. The spectral curve of the general
monopole of this class may be put (by a rotation) in the form
$$\eta^3+\eta(\alpha_0\zeta\sp4
+\alpha_1\zeta\sp3+\alpha\zeta\sp2-\bar\alpha_1\zeta+\bar\alpha_0)
+\beta\zeta\sp6 +\beta_1\zeta\sp5+\beta_2\zeta\sp4+\gamma\zeta\sp3
-\bar\beta_2\zeta\sp2  +\bar\beta_1\zeta   -\beta=0
$$
with $\alpha$, $\beta$ and $\gamma$ real. Hitchin's constraints on
the spectral curve mean there are transcendental relations amongst
these coefficients and the outstanding problem is to realise
these. Two sorts of problem arise. The first is implementing
Hitchin's constraint \textbf{H2} coming from the triviality of the
line bundle $L\sp2$ on the spectral curve. This leads via the
equivalent Ercolani-Sinha constraints (Lemma 2) to requiring the
vector of $\mathfrak{b}$-periods of the meromorphic differential
$\gamma_\infty$ to be a half-period in the period lattice. In
\cite{bren06} it was shown for the trigonal curve
$$\eta^3
+\beta\zeta\sp6 +\beta_1\zeta\sp5+\beta_2\zeta\sp4+\gamma\zeta\sp3
-\bar\beta_2\zeta\sp2  +\bar\beta_1\zeta   -\beta=0
$$
how these might be reexpressed in terms of the four
$\mathfrak{a}$-periods of a specified holomorphic differential.
These constraints were then solved for the symmetric curves
$$\eta^3
+\beta\zeta\sp6 +\gamma\zeta\sp3  -\beta=0,
$$
and a countable family of curves satisfying this constraint of
Hitchin ensued. The problem of requiring a curve with specified
periods of a given meromorphic differential arises in many
settings within finite-gap integration theory. The bijective
correspondence between harmonic maps  $ T ^ 2\rightarrow S ^ 3 $
and algebro-geometric data, specifying curves with given filling
fractions in the AdS/CFT correspondence and seeking closed real
geodesics on an ellipsoid all lead to this problem. Finding ways
to solve such will be an important area for future research.

The second type of problem and the focus of this paper has been in
satisfying Hitchin's constraint \textbf{H3}, the vanishing of a
real one parameter family of cohomologies of certain line bundles,
$H^0(\hat{\mathcal{C}},L^{\lambda}(n-2))=0$ for $\lambda\in(0,2)$.
We reexpressed this in terms of the intersection of a real line
with the theta divisor $\Theta$ of the curve and the problem is to
count the number of intersection points. Again a more general
theory is called for. We made progress here utilizing two features
of the geometry of our situation. The first was that our curve
(\ref{bren03}) has extra symmetry: it falls within a class studied
by Hitchin, Manton and Murray \cite{hmm95} when looking at
spectral curves of monopoles with spatial symmetries. Curves of
the form
\begin{equation}\label{cychmm}
\eta^3+\alpha\eta\zeta^2+\beta\zeta^6+\gamma
\zeta^3-\beta=0
\end{equation} have a cyclic $\texttt{C}_3$ symmetry
$(\zeta,\eta)\rightarrow( \rho\zeta,\rho\eta)$ where $\rho^3=1$.
If $\gamma=0$ this is enlarged to a dihedral symmetry
$\texttt{D}_3$ with $(\zeta,\eta)\rightarrow(
1/\zeta,-\eta/\zeta^2)$. Actually the spectral curves themselves
have larger symmetry (for any $\gamma$ in the curve (\ref{cychmm})
we have the symmetry $(\zeta,\eta)\rightarrow(
-1/\zeta,-\eta/\zeta^2)$) but the nomenclature is based on those
symmetries that may be realized as spatial symmetries. This cyclic
symmetry means there exists a quotient spectral curve. Here it was
of genus 2. We were then able to show that a theorem of Fay and
Accola applied and so the problem reduced to one about the theta
divisor of the quotient curve (Theorem \ref{beH3}). For a
cyclically invariant charge $n$ monopole exactly the same
considerations apply and we find the genus $(n-1)^2$ monopole
curve is an $n$-fold unbranched cover of a genus $(n-1)$
hyperelliptic curve. This is the Affine Toda curve of
Seiberg-Witten theory observed by Sutcliffe \cite{sut96}. Thus
symmetry together with the Fay-Accola theorem reduces the problem
significantly. The second simplifying feature reduced the problem
to one of elliptic functions: Humbert theory tells us our curve
covered an elliptic curve. This feature is a consequence of the
great symmetry of our curve and will not persist for the general
curves (\ref{cychmm}). Notwithstanding this reduction to questions
of elliptic functions we have not proven the general Conjecture
\ref{conjecture} counting the number of intersections of the line
with the theta divisor. We have however established the uniqueness
of the tetrahedrally symmetric monopole within spectral curves of
the form (\ref{bren03}).

An obvious line for further study is to seek those monopoles
within the class (\ref{cychmm}). Hitchin, Manton and Murray argued
that there were five loci of monopoles within this. These loci are
totally geodesic submanifolds of the full moduli space and may be
viewed as the orbits of geodesic monopole scattering. Of these
loci, one corresponded to $\texttt{D}_3$ symmetric monopoles:
asymptotically we have $\alpha^3=27\beta^2$ (with $\beta$ large
and positive at one end and negative at the other) and half-way
along this there is the axisymmetric monopole. The other four loci
were isomorphic: at one end asymptotically one has
$\alpha^3=27\beta^2$ (with $\beta$ of either sign) and $\gamma=0$
while at the other end $\alpha=\pi^2/4-3b^2$, $\beta=0$ and
$\gamma=2b(b^2+\pi^2/4)$ (with $b$ of either sign). Half-way along
this is the tetrahedrally symmetric monopole, the four loci
corresponding to four distinct orientations of the tetrahedron.
Extending our work to this broader class encounters new
difficulties. Although we may use the cyclic symmetry to express
the curve and Ercolani-Sinha constraints to ones for the quotient
curve, the period integrals arising are not simply expressible in
terms of hypergeometric functions and the curve does not cover an
elliptic curve. These are significant complications and we hope to
pursue this elsewhere.

\section*{Acknowledgements}
Both authors are grateful to MISGAM for funding a research visit
of VZE to Edinburgh in 2009.

\appendix
\section{Theta Functions}
 For $r\in \mathbb{N}$ the canonical Riemann
 $\theta$-function is given by
\begin{equation}
\theta(\boldsymbol{z}\,|\,\tau) =\sum_{\boldsymbol{n}\in \mathbb{Z}^r}
\exp(\imath\pi \boldsymbol{n}^T \tau\boldsymbol{n}+2 \imath\pi
\boldsymbol{z}^T \boldsymbol{n}).
\end{equation}
The $\theta$-function is holomorphic on
$\mathbb{C}^r\times\mathbb{S}^r$ and satisfies
\begin{equation}\theta(\boldsymbol{z}+\boldsymbol{p}\,|\,\tau)=
\theta(\boldsymbol{z}\,|\,\tau),\quad
\theta(\boldsymbol{z}+\boldsymbol{p}\tau\,|\,\tau)=
\mathrm{exp}\{-\imath\pi(\boldsymbol{p}^T\tau \boldsymbol{p}
+2\boldsymbol{z}^T\boldsymbol{p})\}\, \theta(\boldsymbol{z}\,|\,\tau),
\label{transformation}
\end{equation}
where $\boldsymbol{p}\in\mathbb{Z}^r$.

The Riemann $\theta$-function
$\theta_{\boldsymbol{a},\boldsymbol{b}}(\boldsymbol{z}\,|\,\tau)$
with characteristics $\boldsymbol{a},\boldsymbol{b}\in\mathbb{Q}$
is defined by
\begin{align*}
\theta_{\boldsymbol{a},\boldsymbol{b}}(\boldsymbol{z}\,|\,\tau)
&=\mathrm{exp} \left\{ \imath\pi
(\boldsymbol{a}^T\tau\boldsymbol{a}
+2\boldsymbol{a}^T(\boldsymbol{z}+\boldsymbol{b})))\right\}
\theta(\boldsymbol{z}+\tau\boldsymbol{a}+\boldsymbol{b}\,|\,\tau)\\
& =\sum_{\boldsymbol{n}\in\mathbb{Z}^r}\mathrm{exp}
\left\{\imath\pi(\boldsymbol{n}+\boldsymbol{a})^T\tau
            (\boldsymbol{n}+\boldsymbol{a})
+2\imath\pi
(\boldsymbol{n}+\boldsymbol{a})^T(\boldsymbol{z}+\boldsymbol{b})
\right\},
\end{align*}
where $\boldsymbol{a},\boldsymbol{b}\in \mathbb{Q}^r$. This is
also written as
$$\theta_{\boldsymbol{a},\boldsymbol{b}}(\boldsymbol{z}\,|\,\tau)=
\theta\left[\begin{matrix}\boldsymbol{a}
\\ \boldsymbol{b}\end{matrix}\right](\boldsymbol{z}\,|\,\tau).
$$
For arbitrary $\boldsymbol{a},\boldsymbol{b}\in \mathbb{Q}^r$ and
$\boldsymbol{a}',\boldsymbol{b}'\in \mathbb{Q}^r$ the following
formula is valid
\begin{align}
\theta_{\boldsymbol{a},\boldsymbol{b}}
(\boldsymbol{z}+\boldsymbol{a}'\tau+\boldsymbol{b}'\,|\,\tau)&=\mathrm{exp}\left\{
-\imath\pi
{\boldsymbol{a}'}^T\tau{\boldsymbol{a}'}-2\imath\pi{\boldsymbol{a}'}^T\boldsymbol{z}
 -2\imath\pi (\boldsymbol{b}+\boldsymbol{b}')^T{\boldsymbol{a}'}   \right\}
 \times
\theta_{\boldsymbol{a}+\boldsymbol{a}',\boldsymbol{b}+\boldsymbol{b}'}
(\boldsymbol{z}\,|\,\tau).\label{thetatransf}
\end{align}

The function $\theta_{\boldsymbol{a},\boldsymbol{b}}(\tau)=
\theta_{\boldsymbol{a},\boldsymbol{b}}(\boldsymbol{0}\,|\,\tau) $
is called the $\theta$-constant with characteristic
$\boldsymbol{a},\boldsymbol{b}$. We have
\begin{align*} \
&\theta_{-\boldsymbol{a},-\boldsymbol{b}}(\boldsymbol{z}\,|\,\tau)=
\theta_{\boldsymbol{a},\boldsymbol{b}}(-\boldsymbol{z}\,|\,\tau)\\
&\theta_{\boldsymbol{a}+\boldsymbol{p},\boldsymbol{b}+\boldsymbol{q}}
(\boldsymbol{z}\,|\,\tau)= \mathrm{exp}(2\pi\imath
\boldsymbol{a}^T\boldsymbol{q})
\theta_{\boldsymbol{a},\boldsymbol{b}}(\boldsymbol{z}\,|\,\tau)
\end{align*}

The following transformation formula is given in \cite[p85,
p176]{igusa72}.
\begin{proposition}
For any $\mathfrak{g}=\left(\begin{array}{cc}A&B\\C&D
\end{array}\right)\in\mathrm{Sp}(2g,\mathbb{Z})$ and
$(\boldsymbol{a},\boldsymbol{b})\in\mathbb{Q}^{2g}$ we put
\begin{align*}
\mathfrak{g}\cdot(\boldsymbol{a},\boldsymbol{b})&=
(\boldsymbol{a},\boldsymbol{b})\mathfrak{g}^{-1}
+\frac12(\mathrm{diag}(CD^T),\mathrm{diag}(AB^T) )\\
\boldsymbol{\phi}_{\boldsymbol{a},\boldsymbol{b}}(\mathfrak{g})&=-\frac12
(\boldsymbol{a}D^TB\boldsymbol{a}^T
-2\boldsymbol{a}B^TC\boldsymbol{b}^T+
\boldsymbol{b}C^TA\boldsymbol{b}^T)
+\frac12(\boldsymbol{a}D^T-\boldsymbol{b}C^T)^T\mathrm{diag}(AB^T)
,\end{align*} where $\mathrm{diag}(A)$ is the row vector
consisting of the diagonal components of $A$. Then for every
$\mathfrak{g}\in\mathrm{Sp}(2g,\mathbb{Z})$ we have
\begin{align}\begin{split}
&\theta_{\mathfrak{g}\cdot(\boldsymbol{a},
\boldsymbol{b})}(0\,|\,(A\tau_\mathfrak{b}+B)(C\tau_\mathfrak{b}+D)^{-1})
=\kappa(\mathfrak{g})\mathrm{exp}(2\pi\imath
\boldsymbol{\phi}_{\boldsymbol{a},\boldsymbol{b}}(\mathfrak{g})   )\,
\mathrm{det}(C\tau_\mathfrak{b}+D)^{\frac12}
\theta_{(\boldsymbol{a},\boldsymbol{b})}(0\,|\,\tau_\mathfrak{b})
\end{split}\label{igusa}
\end{align}
in which $\kappa(\mathfrak{g})^2$ is a $4$-th root of unity
depending only on $\mathfrak{g}$ while
\begin{equation} \begin{split}
\theta_{\mathfrak{g}\cdot(\boldsymbol{a},
\boldsymbol{b})}(z(C\tau_\mathfrak{b}+D)^{-1}\,|\,(A\tau_\mathfrak{b}+B)(C\tau_\mathfrak{b}+D)^{-1})
&=\mu\,\exp\left(i\pi z(C\tau_\mathfrak{b}+D)^{-1}Cz\sp{T}\right)
\,
\mathrm{det}(C\tau_\mathfrak{b}+D)^{\frac12}\\
&\qquad
\times\theta_{(\boldsymbol{a},\boldsymbol{b})}(z\,|\,\tau_\mathfrak{b})
\end{split} \label{igusab}
\end{equation}
and $\mu$ is a complex number independent of $\tau$ and $z$ such
that $|\mu|=1$.
\end{proposition}

\section{Proof of Proposition \ref{humred}}

The proposition follows from Theorem \ref{beH3}.  Let us first
simplify (\ref{tfa}) first using (\ref{modTa}). Using the standard
transformation properties,
\begin{align*}
\vartheta_3\left(z_1\,\Big|\,1- \frac1{\mathcal{T} -2}\right)&=
\vartheta_4\left(z_1\,\Big|\,- \frac1{\mathcal{T} -2}\right) \\
&=\left(-i\left[\mathcal{T} -2\right]\right)\sp{1/2}\,
\exp\left[i\pi z_1\sp2(\mathcal{T} -2)\right]\,
\vartheta_2\left(\left[\mathcal{T} -2\right]z_1\,\Big|\,\mathcal{T}-2\right)\\
&=\left(-i\left[\mathcal{T} -2\right]\right)\sp{1/2}\,
\exp\left[i\pi z_1\sp2(\mathcal{T} -2)-i\pi/2\right]\,
\vartheta_2\left(\left[\mathcal{T} -2\right]z_1\,\Big|\,\mathcal{T}\right)
\end{align*}
\begin{align*}
\vartheta_3&\left(z_1+1/2\,\Big|\,1- \frac1{\mathcal{T} -2}\right)=
\vartheta_4\left(z_1+1/2\,\Big|\,- \frac1{\mathcal{T} -2}\right) \\
&=\left(-i\left[\mathcal{T} -2\right]\right)\sp{1/2}\,
\exp\left[i\pi (z_1+1/2)\sp2(\mathcal{T} -2)\right]\,
\vartheta_2\left(\left[\mathcal{T} -2\right](z_1+1/2)\,\Big|\,\mathcal{T}-2\right)\\
&=\left(-i\left[\mathcal{T} -2\right]\right)\sp{1/2}\,
\exp\left[i\pi z_1\sp2(\mathcal{T} -2)\right]\,
\vartheta_3\left(\left[\mathcal{T} -2\right]z_1\,\Big|\,\mathcal{T}\right)
\end{align*}
\begin{align*}
\vartheta_3\left( 2z_2\,|\,4\widetilde{\tau}_{2,2}\right)&=
\vartheta_3\left( 2z_2\,|\,\frac{\mathcal{T}}{3}-2\right)=
\vartheta_3\left( 2z_2\,|\,\frac{\mathcal{T}}{3}\right)\\
\vartheta_2\left( 2z_2\,|\,4\widetilde{\tau}_{2,2}\right)&=
\vartheta_2\left( 2z_2\,|\,\frac{\mathcal{T}}{3}-2\right)=
\exp(-i\pi/2)\,\vartheta_2\left( 2z_2\,|\,\frac{\mathcal{T}}{3}\right).
\end{align*}
we obtain
\begin{equation}\label{tranfg2t}\begin{split}
\theta(z_1,z_2;\widetilde{\tau})&=
\left(-i\left[\mathcal{T} -2\right]\right)\sp{1/2}\,
\exp\left[i\pi z_1\sp2(\mathcal{T} -2)-i\pi/2\right]\\
&\times\left[
\vartheta_2\left(\left[\mathcal{T} -2\right]z_1\,\Big|\,\mathcal{T}\right)
\vartheta_3\left( 2z_2\,|\,\frac{\mathcal{T}}{3}\right)+
\vartheta_3\left(\left[\mathcal{T} -2\right]z_1\,\Big|\,\mathcal{T}\right)
\vartheta_2\left( 2z_2\,|\,\frac{\mathcal{T}}{3}\right)
\right].
\end{split}
\end{equation}
Now the argument ${\boldsymbol
z}=\lambda\,\boldsymbol{U}'+\boldsymbol{K}' +k\,\boldsymbol{l}'$
is simplified upon setting
$$y=\frac{n+m}{3}\,\rho\,\lambda=\frac{n+m}{3}\,\frac{(-1+i\sqrt{3})}{2}\,\lambda.$$
This yields
$$[\mathcal{T}
-2]z_1=-i\sqrt{3}\,y+\frac43\,\mathcal{T}-3+k-k\,\frac{\mathcal{T}}{3},\qquad
2z_2=y+\frac{\mathcal{T}}6+\frac{k-1}3.
$$
With $A_\tau(v)=e\sp{-i\pi(2v+\tau)}$,
$B_\tau(v)=e\sp{-i\pi(v+\tau/4)}$ we have
\begin{align*}
\vartheta_2\left(\left[\mathcal{T} -2\right]z_1\,\Big|\,\mathcal{T}\right)
&=(-1)\sp{k-1}A_\mathcal{T}(-[i\sqrt{3}\,y+(k-1)\frac{\mathcal{T}}{3}])\,
\vartheta_2\left(i\sqrt{3}\,y+(k-1)\frac{\mathcal{T}}{3}\,\Big|\,\mathcal{T}\right)\\
\vartheta_3\left( 2z_2\,|\,\frac{\mathcal{T}}{3}\right)&=
B_{\mathcal{T}/3}(y+\frac{k-1}3)
\,\vartheta_2\left( y+\frac{k-1}3\,|\,\frac{\mathcal{T}}{3}\right)\\
\vartheta_3\left(\left[\mathcal{T} -2\right]z_1\,\Big|\,\mathcal{T}\right)
&=A_\mathcal{T}(-[i\sqrt{3}\,y+(k-1)\frac{\mathcal{T}}{3}])\,
\vartheta_3\left(i\sqrt{3}\,y+(k-1)\frac{\mathcal{T}}{3}\,\Big|\,\mathcal{T}\right)\\
\vartheta_2\left( 2z_2\,|\,\frac{\mathcal{T}}{3}\right)&=
B_{\mathcal{T}/3}(y+\frac{k-1}3)
\,\vartheta_3\left( y+\frac{k-1}3\,|\,\frac{\mathcal{T}}{3}\right)
\end{align*}
and substituting these in (\ref{tranfg2t}) gives us, up to an
exponential factor, the function $H_{k-1}(y)$ where we define the
functions
\begin{align}
H_k(y):&=
\vartheta_3\left(i\sqrt{3}\,y+\frac{k\mathcal{T}}{3}\,\Big|\,\mathcal{T}\right)
\vartheta_3\left( y+ \frac{k}{3}\,|\,\frac{\mathcal{T}}{3}\right)
+(-1)^{k}\,
\vartheta_2\left(i\sqrt{3}\,y+\frac{k\mathcal{T}}{3}\,\Big|\,\mathcal{T}\right)
\vartheta_2\left( y+ \frac{k}{3}\,|\,\frac{\mathcal{T}}{3}\right),\\
h_{k}(y):&=\frac{H_k(y)}{
\vartheta_2\left(i\sqrt{3}\,y+\frac{k\mathcal{T}}{3}\,\Big|\,\mathcal{T}\right)
\vartheta_3\left( y+ \frac{k}{3}\,|\,\frac{\mathcal{T}}{3}\right)
}\nonumber \\
&=h_{k}^{I}(y)+(-1)^{k}\, h_{k}^{II}(y)=
\frac{\vartheta_3}{\vartheta_2}
\left(i\sqrt{3}\,y+\frac{k\mathcal{T}}{3}\,\Big|\,\mathcal{T}\right)
+(-1)^{k}\, \frac{\vartheta_2}{\vartheta_3}\left( y+ \frac{k}{3}
\,|\,\frac{\mathcal{T}}{3}\right),\label{thetaqn}
\end{align}
where $k\in\mathbb{Z}$ and $\frac{\vartheta_3}{\vartheta_2}
\left(z|\mathcal{T}\right)$ is shorthand for
$\frac{\vartheta_3\left(z|\mathcal{T}\right)}{
\vartheta_3\left(z|\mathcal{T}\right)}$. Thus we have shown
$$f_k(\lambda)=0\Longleftrightarrow H_{k-1}(\lambda)=0.$$
To establish the proposition we observe that the zeros of
$H_{k}(y)$ are different from those of $\vartheta_2\left( y+
\frac{k}{3}\,|\,\frac{\mathcal{T}}{3}\right)$ and
$\vartheta_3\left(i\sqrt{3}\,y+\frac{k\mathcal{T}}{3}\,\Big|\,\mathcal{T}\right)$.
Upon writing $y=w(-1+i\sqrt{3})$ with $w$ real a zero of
$\vartheta_2\left( y+
\frac{k}{3}\,|\,\frac{\mathcal{T}}{3}\right)$ takes the form
$$y=w(-1+i\sqrt{3})=m_1+\frac12-\frac{k}3+n_1\,\frac{\mathcal{T}}{3},\qquad n_1,m_1\in\mathbb{Z}$$
while a zero of the latter has the form
$$i\sqrt{3}\, y=w(-i\sqrt{3}-3)=m_2+\frac12+(n_2+\frac12-\frac{k}3)\,\mathcal{T},
\qquad n_2,m_2\in\mathbb{Z}.$$
If these were to vanish simultaneously then we see from their
imaginary parts that
$$0=n_2+\frac{n_1}3+\frac12-\frac{k}3,$$
which is not possible. Hence the zero's of $H_k(y)$ are different
from those of $\vartheta_2\left( y+
\frac{k}{3}\,|\,\frac{\mathcal{T}}{3}\right)$ and
$\vartheta_3\left(i\sqrt{3}\,y+\frac{k\mathcal{T}}{3}\,\Big|\,\mathcal{T}\right)$.
Therefore
$$f_k(\lambda)=0\Longleftrightarrow H_{k-1}(y(\lambda))=0\Longleftrightarrow h_{k-1}(y(\lambda))=0
$$ and
(\ref{vanishing11a}) is established up making use of Theorem
\ref{beH3}.

The first periodicity of (\ref{hper}) is follows immediately from
the periodicity of the theta functions. For the second we note
\begin{align*}
h_{k}^{I}\left(y+2\frac{(n+m)}3\right)&=
\frac{\vartheta_3}{\vartheta_2}\left(i\sqrt{3}\, y+\frac{(2n-m)}3\mathcal{T}+\frac{k\mathcal{T}}3
)\,\Big|\,\mathcal{T}\right)
=h_{k-[n+m]}^{I}(y),\\
h_{k}^{II}\left(y+2\frac{(n+m)}3\right)&
=(-1)\sp{n+m}\,h_{k-[n+m]}^{II}(y),\\
h_{k}\left(y+2\frac{(n+m)}3\right)&
=h_{k-[n+m]}(y),\\
h_{k}^{I}\left(y+2\frac{(n+m)}3\rho\right)&=
\frac{\vartheta_3}{\vartheta_2}\left(i\sqrt{3}\, y-\frac{(n+m)}3i\sqrt{3}-(n+m)+\frac{k\mathcal{T}}3
)\,\Big|\,\mathcal{T}\right)\\
&=(-1)\sp{n+m}\,
\frac{\vartheta_3}{\vartheta_2}\left(i\sqrt{3}\, y-\frac{(2n-m)}6\mathcal{T}+\frac{k\mathcal{T}}3
)\,\Big|\,\mathcal{T}\right)\\
&=(-1)\sp{n+m}\,\left[h_{k-[n+m]}\sp{I}(y)\right]\sp{\epsilon(m)}\\
h_{k}^{II}\left(y+2\frac{(n+m)}3\rho\right)&
=\frac{\vartheta_2}{\vartheta_3}\left(y+\frac{k-[n+m]}3+\frac{(2n-m)}6\mathcal{T}
\,\Big|\,\frac{\mathcal{T}}{3}\right)
=\left[
h_{k-[n+m]}^{II}(y)
\right]\sp{\epsilon(m)}\\
h_{k}\left(y+2\frac{(n+m)}3\rho\right)
&=\begin{cases}
(-1)\sp{n+m}\,h_{k-[n+m]}(y)&\text{if }m\text{ even},\\ \\
(-1)\sp{k}g_{k-[n+m]}(y)
&\text{if }m\text{ odd},
\end{cases}
\end{align*}
where $\epsilon(m)=1$ if $m$ is even and $-1$ if $m$ is odd and
$$
g_{k}(y):=h_k\left(y+\frac{\mathcal{T}}2\right)=\frac{H_k(y)}{
\vartheta_3\left(i\sqrt{3}\,y+\frac{k\mathcal{T}}{3}\,\Big|\,\mathcal{T}\right)
\vartheta_2\left( y+ \frac{k}{3}\,|\,\frac{\mathcal{T}}{3}\right)
}$$
Now $y(\lambda+2)=y(\lambda)+2(n+m)\rho/3$ and the result follows
as $h_k(y)=0\Longleftrightarrow g_{k}(y)=0$.

\section{Proof of Proposition \ref{formbmn}}

The proof involves three steps. First let us parameterize
$M=(2\imath -b)^{\frac13}/(b^2+4)^{\frac16}$ of Lemma
\ref{covering} by
\begin{equation}M=\frac{p+\imath\sqrt{3}}{p-\imath\sqrt{3}}.\label{subm} \end{equation}
Then solving for $b$ we obtain the form (\ref{bformula}),
\begin{equation}
b=-\frac{\sqrt{3}(p^6-45p^4+135p^2-27) }{9p(p^4-10p^2+9)}.
\label{subbp} \end{equation}
The same substitution (\ref{subm}) in
the Jacobian moduli (\ref{jacobimodpm}) leads to their
parametrization
\begin{equation}
k_+^2=\frac{(p+1)^3(3-p)}{16p},\quad k_-^2=\frac{(p+1)(3-p)^3}{16p^3}. \label{rampar}
\end{equation}
This is the parametrization used by Ramanujan in his
hypergeometric relations of signature 3, see e.g. [BBG95]. The
$\theta$-functional representation of the moduli $k_{\pm}$ can be
found in [Law89, Section 9.7],
\begin{equation}
k_+=\frac{\vartheta_2^2(0|\tau)}{\vartheta_3^2(0|\tau)},\quad
k_-=\frac{\vartheta_2^2(0|3\tau)}{\vartheta_3^2(0|3\tau)}, \quad
p= \frac{3\vartheta_3^2(0|3\tau)}{\vartheta_3^2(0|\tau)} .
\label{ramparb}
\end{equation}

To establish the proposition we must show that the parameter
$\tau=\mathcal{T}(m,n)/6$. To do this we next consider a genus two
curve with period system $\tau_{11},\frac12,\tau_{22}$. Bolza
\cite[p.451]{Bol886} showed that the associated genus two curve
may be presented in the form
\begin{equation} \label{bolzacurve}
w^2=(1+z^2)(c^2+{e'}^2z^2)({c'}^2+{e}^2z^2)
\end{equation}
with $c=\theta_2^2/\theta_3^2$, $c'=\theta_4^2/\theta_3^2$,
$e=\Theta_2^2/\Theta_3^2$, $e'=\Theta_4^2/\Theta_3^2$ and
 $\theta_j=\vartheta_j(0\vert 2\tau_{11}),\; \Theta_j=\vartheta_j(0\vert 2\tau_{22}),\; j=2,3,4. $
For the moment let us assume this is our curve $\mathcal{C}$. Both
the (unnormalized) holomorphic differentials $z\mathrm{d}z/w$ and
$\mathrm{d}z/w $ reduce to the holomorphic differentials of
elliptic curves that we shall call (and shortly identify with)
$\mathcal{E}_{\pm}$ by the (respective) substitutions $z^2=t$ and
$z^2=1/t$,
\begin{equation}
\frac12\frac{\mathrm{d}t}{\sqrt{(1+t)(c^2+{e'}^2t)({c'}^2+{e}^2t)}},
\quad -\frac12\frac{\mathrm{d}t}{\sqrt{(1+t)(c^2t+{e'}^2)({c'}^2t+{e}^2)}}.
\end{equation}
The Jacobian moduli of $\mathcal{E}_{\pm}$ are then
\begin{equation}  k_+^2=\frac{{e}^2}{{e'}^2}\frac{c^2-{e'}^2}{{c'}^2-e^2},
\quad k_-^2=\frac{{c'}^2}{{c}^2}\frac{c^2-{e'}^2}{{c'}^2-e^2}\quad
.\label{kmoduli} \end{equation} Taking the period system (4.14), i.e.
$\tau_{11}=-1/(\mathcal{T}+6)$, $\tau_{22}=(\mathcal{T}+6)/12$,
one can see that
\[ c=\frac{1}{C'},\; c'=\imath \frac{C}{C'},\; e=-\imath\frac{E}{E'},\;e'=\frac{1}{E'} \]
where
\[ C=\frac{\vartheta_2^2}{\vartheta_3^2}\left(0\vert \frac{\mathcal{T}}{6}  \right),\;
C'=\frac{\vartheta_4^2}{\vartheta_3^2}\left(0\vert
\frac{\mathcal{T}}{6}  \right),
E=\frac{\vartheta_2^2}{\vartheta_3^2}\left(0\vert
\frac{\mathcal{T}}{2}  \right),\;
E'=\frac{\vartheta_4^2}{\vartheta_3^2}\left(0\vert
\frac{\mathcal{T}}{2}  \right). \] Again using the parametrization
(\ref{rampar}, \ref{ramparb}) we obtain from (\ref{kmoduli}) after
simplification,
\begin{align*}
k_+^2=C^2=\frac{\vartheta_2^4}{\vartheta_3^4}
\left(0|\frac{\mathcal{T}}{6}\right),\quad k_-^2=E^2=\frac{\vartheta_2^4}{\vartheta_3^4}
\left(0|\frac{\mathcal{T}}{2}\right).
\end{align*}
This has proven $\tau=\mathcal{T}(m,n)/6$ provided we can
establish the curve (\ref{bolzacurve}) is $\mathcal{C}$ and
identify the elliptic curves $\mathcal{E}_\pm$ as above.

Our final step then is to show the curve  (\ref{bolzacurve}) is
birationally equivalent to the curve (\ref{hcurve}). Indeed let
$$T=\frac{L+\mu}{L-\mu},\ S=\frac{8\nu}{(L-\mu)\sp3},\qquad
\mu=L\,\frac{T-1}{T+1},\ \nu=\frac{L^3 S}{(T+1)^3}.$$ Then
(\ref{hcurve}) transforms to
$$S^2=(T-1)^6+2\frac{b}{L^3}(T^2-1)^3+(T+1)^6.$$
This curve is of the same form as (\ref{bolzacurve}) up to scaling
of $S$ and $T$. Using the substitution (\ref{subbp}) this becomes
\begin{align*}
\left(\frac{S}{\sqrt{2(1-b/L^3)}}\right)\sp2&=
27\,{\frac { \left( p-1 \right) ^{2} \left( p+1 \right) ^{2}}{{p}^{2}
 \left( p+3 \right) ^{2} \left( -3+p \right) ^{2}}}\,T^6+
 9\,{\frac {45\,{p}^{2}+18+{p}^{6}}{{p}^{2} \left( p+3 \right) ^{2}
 \left( -3+p \right) ^{2}}}\,T^4\\
&\qquad + 3\,{\frac {2\,{p}^{6}+45\,{p}^{4}+81}{{p}^{2} \left( p+3 \right) ^{2}
 \left( -3+p \right) ^{2}}}\,T^2
+1
\intertext{whereas (\ref{bolzacurve}) may be written}
\left(\frac{w}{cc'}\right)\sp2&=
{\frac { \left( p+3 \right) ^{4} \left( p+1 \right) ^{2}}{ \left( p-1
 \right) ^{4}{p}^{2} \left( -3+p \right) ^{2}}}\,z^6+
 {\frac { \left( 45\,{p}^{2}+18+{p}^{6} \right)  \left( p+3 \right) ^{2
}}{ \left( p-1 \right) ^{4}{p}^{2} \left( -3+p \right) ^{2}}}\,z^4\\
&\qquad+
{\frac {2\,{p}^{6}+45\,{p}^{4}+81}{{p}^{2} \left( -3+p \right) ^{2}
 \left( p-1 \right) ^{2}}}
\,z^2+1.
\end{align*}
These coincide with $z= \sqrt{3}\,\left( p-1 \right)T/ \left( p+3
\right)$. The substitution $W=T^2$ reduces the canonical
differentials ${d}T/S$ and $T\,{d}T/S$ to the canonical
differentials of the elliptic curves $\mathcal{E}_\pm$ given in
Lemma \ref{covering}. These correspond to the differentials and
curves $\mathcal{E}_\pm$ identified above.

In the course of the proof we find the $\theta$-constant
representation for the Jacobian moduli of the curves
$\mathcal{E}_{\pm}$,
\begin{equation}
k_+(m,n)=\frac{\vartheta_{2}^2\left(0\vert \frac{\mathcal{T}(m,n)}{6}\right)}
{\vartheta_{3}^2\left(0\vert \frac{\mathcal{T}(m,n)}{6}\right)},
\quad k_-(m,n)=\frac{\vartheta_{2}^2\left(0\vert \frac{\mathcal{T}(m,n)}{2}\right)}
{\vartheta_{3}^2\left(0\vert \frac{\mathcal{T}(m,n)}{2}\right)},
\quad \mathcal{T}(m,n)=2\imath\sqrt{3}\frac{n+m}{2n-m}
\end{equation}
which yields the corollary.

\providecommand{\bysame}{\leavevmode\hbox
to3em{\hrulefill}\thinspace}
\bibliographystyle{amsalpha}

\end{document}